# $k$-space Physics-informed Neural Network (k-PINN) for Compressed Spectral Mapping and Efficient Inversion of Vibrations in Thin Composite Laminates

Saeid Hedayatrasa [1,2*], Olga Fink [3*], Wim Van Paepegem [1] and Mathias Kersemans [1*]

[1] Mechanics of Materials and Structures (UGent-MMS), Department of Materials, Textiles and Chemical Engineering (MaTCh), Ghent University, Technologiepark-Zwijnaarde 46, 9052 Zwijnaarde, Belgium

[2] Flanders Make-MotionS, 3920 Lommel, Belgium

[3] Intelligent Maintenance and Operating Systems (IMOS) lab, EPFL, Switzerland

* Corresponding authors:
saeid.hedayatrasa@flandersmake.be, olga.fink@epfl.ch, mathias.kersemans@ugent.be

**Abstract**

The vibrational response of structural components carries valuable information about their underlying mechanical properties, health status and operational conditions. This underscores the need for the development of efficient physics-based inversion algorithms which, given a limited set of sensing data points and in the presence of measurement noise, can reconstruct the response at locations where measurement data is not available and/or identify the unknown mechanical properties. Addressing this challenge, Physics-Informed Neural Networks (PINNs) have emerged as a promising approach. PINNs seamlessly integrate governing equations into their architecture and have gained significant interest in solving inversion problems. In the context of learning and inversion of multimodal, multiscale vibrational responses, this paper introduces a novel spectral extension of PINNs, utilizing Fourier basis functions in the wavenumber domain, commonly known as $k$-space. The proposed method, referred to as k-space PINN (k-PINN), offers a robust framework for adjusting complexity and wavenumber composition of the response. Notably, the spectral formulation of k-PINN, coupled with the generally sparse representation of vibrations in k-space, facilitate efficient reconstruction and learning of broadband vibrations and alleviate the spectral bias associated with standard PINN. Additionally, the spectral solution space introduced by k-PINN substantially reduces the computational cost associated with computing physics-informed loss terms. We evaluate the effectiveness of the proposed methodology on reconstructing the bending vibrational mode shapes of a thin composite laminate and identifying its effective bending stiffness coefficients. Mode shapes are initially obtained from finite element simulation, and virtual test data with added noise are generated for evaluation purposes. It is shown that the proposed k-PINN methodology outperforms the standard PINN in terms of both learning and computational efficiency. The performance of k-PINN is further demonstrated by show casing its capability in learning different selections of symmetric, anti-symmetric and asymmetric mode shapes.

## 1. Introduction

Analysis of the dynamic behaviour of solid materials and structural components has been an indispensable approach for evaluating their mechanical properties, integrity and safe operation, i.e. non-destructive testing and structural health monitoring. Dynamic loads may be externally applied (e.g. excitation of a test coupon with piezoelectric actuators) or may be intrinsic to the operational conditions (e.g. interaction of an air foil with aerodynamic forces). Such dynamic perturbation leads to the propagation of elastic waves throughout the material, globally deform the test piece and stimulate its vibrational mode shapes. The measured response then reflects the effective mechanical properties, and the integrity of the test piece (e.g. security of the connections and joints, and presence of internal damage). The response may be measured contact-free, for example, using a scanning laser Doppler vibrometer [1–3], or by deploying a number of sensors, such as piezoelectric transducers [4; 5]. As such, considering a dense measurement grid could lead to tedious scanning or costly utilization of numerous sensors. Moreover, there might be regions with limited access for direct measurement of the response, for example, because embedding a sensor would be impractical or would compromise the integrity of the component. Hence, it is of great importance to develop efficient physics-based inversion algorithms to identify the underlying material properties, and to reconstruct the response at inaccessible or unmeasured regions, from fewer, and possibly sub-wavelength, measurement data in the presence of measurement noise.

There are different inversion algorithms available for this purpose, each of which has its own merits. One may couple a simulation model with an optimization algorithm for model-based inversion of measured data, which requires numerous high-fidelity forward modelling of the problem until convergence is achieved [6–8]. Another alternative, to overcome the computational burden, would be online inversion of measurement data by the offline training of a deep neural network (NN) using simulation/experimental data [2; 9]. Despite its fast online deployment, this technique is only reliable as long as the training dataset is sufficiently large such that it encompasses a wide range of scenarios expected in practice and is representative of the current operational conditions. For unseen cases falling outside the training dataset, it fails to generalize and acts as a 'black box', lacking any interpretability. Compressed sensing technique may also be applied to recover the high-resolution spatial response from sparse sub-wavelength measurement data [10; 11]. Compressed sensing leverages the spectral sparsity of the vibrational response (in the wavenumber- and frequency-domain), estimating the response as a linear superposition of a set of basis components (sensing matrix). The sensing matrix may be based on an analytical closed form solution of the problem, for reconstruction of the response at any given location, with material properties as a priori knowledge [12; 13].

Physics-informed neural networks (PINNs) [14; 15], and more broadly physics-informed machine learning [16], represent a new paradigm in the field of scientific machine learning, which has shown great promise in solving complex problems in science and engineering e.g. [17–21]. The power of PINNs lies in their ability to incorporate the underlying physical laws and inherent knowledge about the problem into the machine learning architecture and its learning process. This leads to a higher reliability and generalizability of the solutions provided by PINNs, even with often limited measurement data. The integration of physics-based principles with data-driven insights through PINNs offers enhanced capabilities for solving inverse problems, making PINNs an invaluable tool [22–24]. This hybrid approach leverages the strengths of both fields, ensuring predictions are accurate, reliable, and interpretable, while adhering to physical laws and reducing reliance on extensive labelled data. This expands their utility and effectiveness in complex problem-solving scenarios. It has been shown that PINNs can even further correct the model misspecification, given sparse noisy data [25].

In this regard, there has been particular interest in applying PINNs in non-destructive testing and structural health monitoring, to identify unknown material properties and/or reconstruct internal features or voids, from potentially sparse measurement data captured at the surface of a specimen [23; 26–28]. One of the challenges in implementing PINNs for the spatial-temporal reconstruction of vibrational responses is the spectral bias inherited from its NN learning basis. It is well known that deep, fully connected NNs tend to learn the low frequency (i.e. less complex) components of the response function, and may (or may not) exploit the higher frequency (i.e. more complex) components in the later stages of training [29]. This has been theoretically proven by constructing the neural tangent kernel matrix of (PI)NNs and observing that higher frequency Eigen functions of NNs are associated with lower eigenvalues and, consequently, a lower learning rate [30]. As a result, learning high frequency or broadband (multiscale) response functions requires particular attention in designing the NN architecture and tuning its hyperparameters. The convergence to multiscale, high-frequency features can be improved by properly initializing the scaling hyper-parameters and further implementing adaptive scaling parameters in the activation functions [31; 32]. Fourier (sine and cosine) basis functions can also be deterministically incorporated into the NN input layer, over a given frequency bandwidth, using the so-called Fourier feature mapping method [33–35]. Due to the presence of nonlinear activation functions, these techniques do not maintain an interpretable relationship between the NN's spectral hyper-parameters and the obtained response function.

The performance of deep learning algorithms in the wavenumber domain ($k$-space) for accelerated magnetic resonant imaging (MRI) has been shown to outperform traditional image-domain approaches [36; 37]. Furthermore, the efficiency of data processing in $k$-space, especially for wavefield analysis and damage detection, is well-established [1; 38; 39]. Converting spatial data to k-space using Fourier transform decomposes it to a set of real and imaginary coefficient maps. These maps reveal the wavenumber composition, highlighting spatial features, their length scale and directionality. The $k$-space representation of data is generally sparse, meaning only a limited number of spectral components are needed to reconstruct the original spatial data. This $k$-space sparsity is in fact a key enabler of the compressed sensing approach [10; 11].

In this research, we propose a $k$-space variant of PINNs for compressed spectral mapping of vibrations, holding interpretable information about the wavenumber composition and symmetry of the response function. The proposed $k$-space PINN (k-PINN) employs Fourier basis functions to construct the response function and learns the relevant real and imaginary coefficient maps. This way, k-PINN establishes a rigorous domain for adjusting the complexity and wavenumber composition of the response and mitigates the spectral bias of PINNs. A proper centralization of the spatial domain further enables decoupling of the symmetric (anti-symmetric) components of the response function through real (imaginary) coefficients. The measured data, depending on their spatial resolution, can inform about the spectral composition of the response. Moreover, considering the spectral sparsity of the response, the underlying physics equations can be more efficiently reconstructed from a finite number of components. The broadband spectral formulation of k-PINN, combined with the generally sparse representation of vibrations in $k$-space, enable efficient learning and reconstruction of multiple vibrational modes. Additionally, the spectral definition of solution space, significantly reduces the computational cost of computing physics-informed loss terms.

We evaluate the performance and the capabilities of the proposed methodology on the case study of bending vibrations in a thin fibre reinforced composite laminate with homogenized orthotropic elastic behaviour. Fiber reinforced composites have a large design freedom and high mechanical performance (e.g. high specific stiffness and corrosion resistance), making them attractive for use in primary and secondary (thin-walled) structural components, signifying the need for their vibrational assessment. We evaluate k-PINN against the standard PINN for this vibrational case study, and demonstrate its performance in the full-field reconstruction of different mode shapes and identification of the effective bending stiffness coefficients, using different numbers and spatial distributions of the data points.

The key novelties and contributions of this paper can be summarized as follows:

- Incorporating a k-space solution space in the PINN architecture for adjusting the spectral composition of the response function, mitigating the spectral bias, reducing the computational cost of PINN, and obtaining a compressed and interpretable representation of the response.
- Adaptive regularization of the k-space using a Gaussian mask for efficient and reliable multimodal reconstruction of vibrational responses.
- Introducing a customized physics-informed output activation function for the efficient reconstruction of the second order derivative terms of the governing equation, which is essential for successfully solving the forth order equation of bending vibrations in thin composite laminates

The remainder of the paper is organized as follows: Section 2 summarizes the vibrational bending behavior of a thin composite laminate in both real- and $k$-space representation, and generates a set of virtual test data with added noise using finite element (FE) simulation. Section 3 introduces the proposed k-PINN architecture and its specialized implementation for the vibrational case study in detail. In section 4, the standard PINN and the proposed k-PINN are applied to this vibrational case study, and the results are presented and discussed.

## 2. Bending Vibrations of a Thin Composite Laminate

In this section, first the governing equations of the bending vibrations of thin composite laminates are summarized. Then the vibrational case study is introduced, and its FE simulation for vibrational modal analysis and the generation of virtual tests data, are explained. Lastly, the $k$-space definition of mode shapes is formulated, and selected mode shapes and their $k$-space coefficient maps are demonstrated.

### 2.1. Governing Equations

By considering a laminate with a symmetric cross-ply layup, the bending-twisting and bending-stretching couplings are absent, and according to classic laminate theory, the constitutive bending stress-strain equations are as follows [40]:

$$\boldsymbol{M} = \begin{Bmatrix} M_{xx} \\ M_{yy} \\ M_{xy} \end{Bmatrix} = \begin{bmatrix} D_{11} & D_{12} & 0 \\ D_{12} & D_{22} & 0 \\ 0 & 0 & D_{66} \end{bmatrix} \begin{Bmatrix} \gamma_{xx} \\ \gamma_{yy} \\ \gamma_{xy} \end{Bmatrix} \tag{1}$$

where $M_{xx}, M_{yy}$ and $M_{xy}$ represent the bending moments, $D_{ij}$ are the effective bending stiffness coefficients derived from the ply's thickness $h_p$ and its orthotropic engineering elastic constants $E_{11}, E_{22}, G_{12}$ and $\nu_{12}$ (see the Appendix), and $\gamma_{xx}, \gamma_{yy}$ and $\gamma_{xy}$ are the bending curvatures. In the case of thin laminates with zero transversal shear strain, these curvatures reduce to:

$$\{\gamma_{xx} \quad \gamma_{yy} \quad \gamma_{xy}\} = \left\{-\frac{\partial^2 w}{\partial x^2} \quad -\frac{\partial^2 w}{\partial y^2} \quad -2\frac{\partial^2 w}{\partial x\, \partial y}\right\} \tag{2}$$

Herein, $w$ is the bending deflection, which is constant through the thickness and the sole degree of freedom at each point on the $xy$-plane. In the absence of extensional modes, the in-plane displacement components at a distance $z$ from the mid-plane are:

$$\{u \quad v\} = \left\{-z\frac{\partial\, w}{\partial x} \quad -z\frac{\partial\, w}{\partial y}\right\} \tag{3}$$

Considering a lateral pressure $P$ (along z-axis), the equilibrium of vertical inertial and elastic forces gives:

$$I_d \frac{\partial^2 \overline{w}}{\partial t^2} - \left(\frac{\partial Q_x}{\partial x} + \frac{\partial Q_y}{\partial y}\right) - P = 0 \tag{4}$$

and according to the equilibrium of the bending moments and rotary inertia, the bending shear tractions are derived as follows:

$$\boldsymbol{Q} = \begin{Bmatrix} Q_x \\ Q_y \end{Bmatrix} = \begin{Bmatrix} \dfrac{\partial M_{xx}}{\partial x} + \dfrac{\partial M_{yx}}{\partial y} + I_r \omega^2 \dfrac{\partial^2 w}{\partial x^2} \\ \dfrac{\partial M_{yy}}{\partial y} + \dfrac{\partial M_{yx}}{\partial x} + I_r \omega^2 \dfrac{\partial^2 w}{\partial y^2} \end{Bmatrix} \tag{5}$$

where $I_d$ and $I_r$ are the deflection and rotary moments of inertia per unit length, respectively. By considering a laminate comprising $N_p$ identical constitutive plies, these can be defined as:

$$I_d = N_p h_p \rho, I_r = \frac{1}{12} \rho \left(N_p h_p\right)^3 \tag{6}$$

Subsequently, the governing equation of motion is derived by substituting Equation 5 into Equation 4. With the assumption of periodic vibration at a circular frequency $\omega = 2\pi f$, this is reduced to:

$$D_{11} \frac{\partial^4 w}{\partial x^4} + 2(D_{12} + 2D_{66}) \frac{\partial^4 w}{\partial x^2 \partial y^2} + D_{22} \frac{\partial^4 w}{\partial y^4} - I_d \omega^2 w + I_r \omega^2 \left(\frac{\partial^2 w}{\partial x^2} + \frac{\partial^2 w}{\partial y^2}\right) - P = 0 \tag{7}$$

At the modal frequencies, and under the assumption of negligible material damping, the governing Equation 7 holds in the absence of any external load $(\mathrm{i.e.} P = 0)$. In fact, the purpose of lateral excitation is solely to stimulate the mode shapes of the plate and to compensate for the energy dissipation caused by damping.

The natural boundary conditions effective at the free edges are:

$$\begin{cases} M_{nn} = 0 \\ V_n = Q_n + \dfrac{\partial M_{ns}}{\partial s} = 0 \end{cases} \tag{8}$$

where $n$ and s indicate the axes normal and tangential to the edge, respectively. $V_n = 0$ is the Kirchhoff free edge condition, which is known to be more accurate than individually setting the shear vertical load $Q_n$ and the twisting moment $M_{ns}$ to zero. This approach compensates for the assumed zero transversal shear deformation of thin plates [23]. Considering Equations 1 and 5, the Kirchhoff free edge condition of the rectangular plate can be derived as:

$$\begin{Bmatrix} V_x \\ V_y \end{Bmatrix} = \begin{Bmatrix} -D_{11} \dfrac{\partial^3 w}{\partial x^3} - (D_{12} + 4D_{66}) \dfrac{\partial^3 w}{\partial x \, \partial y^2} + I_r \omega^2 \dfrac{\partial^2 w}{\partial x^2} \\ -D_{22} \dfrac{\partial^3 w}{\partial y^3} - (D_{12} + 4D_{66}) \dfrac{\partial^3 w}{\partial x^2 \, \partial y} + I_r \omega^2 \dfrac{\partial^2 w}{\partial y^2} \end{Bmatrix} = \begin{Bmatrix} 0 \\ 0 \end{Bmatrix} \tag{9}$$

### 2.2. FE Simulation and Virtual Test Data

Virtual test data were generated for a rectangular carbon fibre reinforced polymer (CFRP) composite laminate with a symmetric layup $[(0/90)_2]_s$, dimensions $750 \times 500\ \mathrm{mm}^2$, and a total thickness of $h = 2\ \mathrm{mm}$, as illustrated in Figure 1. Symmetric boundary conditions (BCs), i.e., a plate supported at the four corners, are considered to study the reconstruction of mode shapes with rectangular symmetry and anti-symmetry. Asymmetric BCs, i.e., a faulty/free bottom-right corner, are additionally considered to study the reconstruction of asymmetric mode shapes.

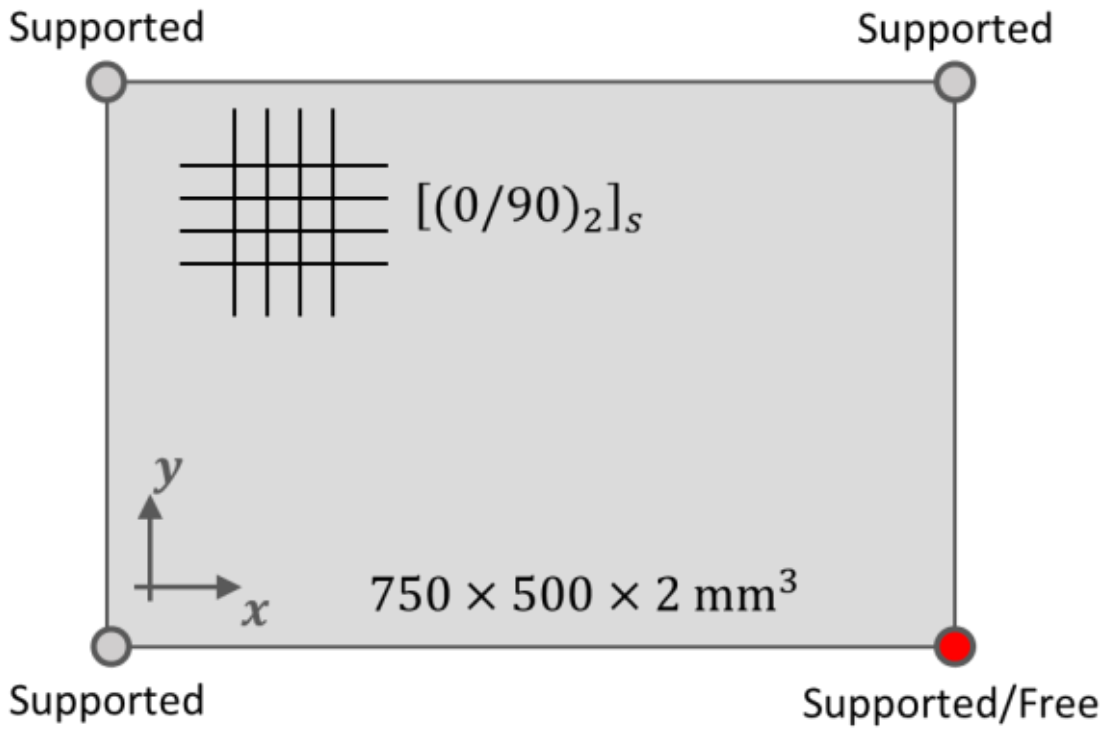

Figure 1. Simulated CFRP laminate simply supported at the corners, representing symmetric boundary conditions (BCs) , and including the possibility of a free bottom-right corner to illustrate asymmetric BCs (e.g. induced by a loose joint).

The CFRP plate was simulated using the Abaqus FE solver, and a modal analysis was performed to obtain the modal behaviour up to a frequency of $f = 1000\ \mathrm{Hz}$. The plate was uniformly meshed with bi-linear triangular shell elements S3, employing a thin element formulation, with an element width of 1 mm, and incorporating a composite section. The in-plane elastic properties of the unidirectional CFRP ply, and effective bending stiffness components corresponding to the simulated laminate with a cross-ply layup $[(0/90)_2]_s$, are listed in Table 1.

Table 1. Assumed in-plane elastic ply properties, and corresponding bending stiffness coefficients for the simulated laminate with cross-ply layup $[(0/90)_2]_s$

| Material | Elastic Constants | | | | Bending Stiffness $[\mathrm{N.m}]$ | | | |
|---|---|---|---|---|---|---|---|---|
| | $E_{11}\,[\mathrm{GPa}]$ | $E_{22}\,[\mathrm{GPa}]$ | $G_{12}\ [\mathrm{GPa}]$ | $\nu_{12}$ | $D_{11}$ | $D_{22}$ | $D_{12}$ | $D_{66}$ |
| CFRP [41] | 108.87 | 9.61 | 5.11 | 0.37 | 52.54 | 27.42 | 2.40 | 3.41 |

In the remainder of this study, the simulated results are contaminated with additive Gaussian white noise and used as virtual test data, $w_{\mathrm{T}}$:

$$w_{\mathrm{T}} = w_{\mathrm{FEA}} + \varepsilon \tag{10}$$

where $\varepsilon$ represents zero-mean Gaussian white noise, regenerated and scaled for each individual mode according to its maximum magnitude to achieve a specified signal-to-noise ratio (SNR) as follows:

$$SNR(m) = 20\log_{10}\left(\frac{\max_i\left(w_{\text{FEA}}(i,m)\right)}{\sqrt{\frac{1}{N}\sum_{i=1}^{N}\varepsilon^2(i,m)}}\right) \tag{11}$$

In this study, a constant SNR of 20dB is considered, which means that the maximum deflection of each mode shape is 10 times greater than the noise floor.

### 2.3. Vibrational Mode shapes in $\boldsymbol{k}$-space

The bending vibrational mode shape of the plate, represented on a uniform grid of $N_x \times N_y$ data points in its $xy$-plane, can be transformed into $k$-space using a 2D discrete Fourier transform (DFT) as follows:

$$\begin{aligned}\tilde{w}(k_x,k_y,f) &= \tilde{w}_{\text{Re}}(k_x,k_y,f) + i\tilde{w}_{\text{Im}}(k_x,k_y,f)\\ &= \frac{1}{N_x \times N_y}\sum_{n_x=0}^{N_x-1}\sum_{n_y=0}^{N_y-1} w\left(x(n_x),y(n_y),f\right)\exp\Big(-2\pi i\big(k_x x(n_x)\\ &\quad + k_y y(n_y)\big)\Big)\end{aligned} \tag{12}$$

which projects the response to the wave vector:

$$\begin{aligned}\boldsymbol{k}(m_x,m_y) &= \{k_x(m_x) \quad k_y(m_y)\}\\ &= \left\{\frac{m_x}{N_f L_x} \quad \frac{m_y}{N_f L_y}\right\}, \begin{cases} m_x = -N_f \times B_x, \dots, 0, 1, 2, \dots, N_f \times B_x \\ m_y = -N_f \times B_y, \dots, 0, 1, 2, \dots, N_f \times B_y\end{cases}\end{aligned} \tag{13}$$

where $B_x$ and $B_y$ determine the bandwidth over which the mode shape is approximated and periodized in the spatial domain, and $N_f \geq 1$ is an integer for increasing the frequency resolution by including fractional harmonics. According to the Nyquist theorem, the bandwidth cannot exceed half of the sampling frequency, i.e. $\{B_x, B_y\}_{Max} = \{0.5(N_x - 1), 0.5(N_y - 1)\}$. Given a uniform grid of data, DFT can be efficiently calculated through the fast Fourier transform (FFT). Figure 2 shows selected mode shapes of the plate and corresponding $k$-space coefficient images calculated through 2D-FFT. A uniform grid of bending deflection data with a spatial resolution of $60 \times 40$ is used, resulting in a bandwidth of $\{B_x, B_y\} = \{29,19\}$, and with no fractional harmonics (i.e. $N_f = 1$).

The selected mode shapes include a symmetric mode S25 (Figure 2(a-c)) and an anti-symmetric mode A24 (Figure 2(d-f)), corresponding to the plate with symmetric boundary conditions (BCs) supported at the four corners, as well as an asymmetric mode shape AS25 (Figure 2(g-i)) corresponding to the plate with asymmetric BCs, i.e., a free bottom-right corner. It can be observed that the symmetric mode S25 is defined purely by real components, the anti-symmetric mode A24 is predominantly defined by the imaginary components, and the asymmetric mode AS25 features a comparable contribution of both real and imaginary components. The results further confirm that the quadrants of $k$-space are diagonally dependent and complex conjugates, indicating that the entire $k$-space is redundant for the reconstruction of the mode shapes.

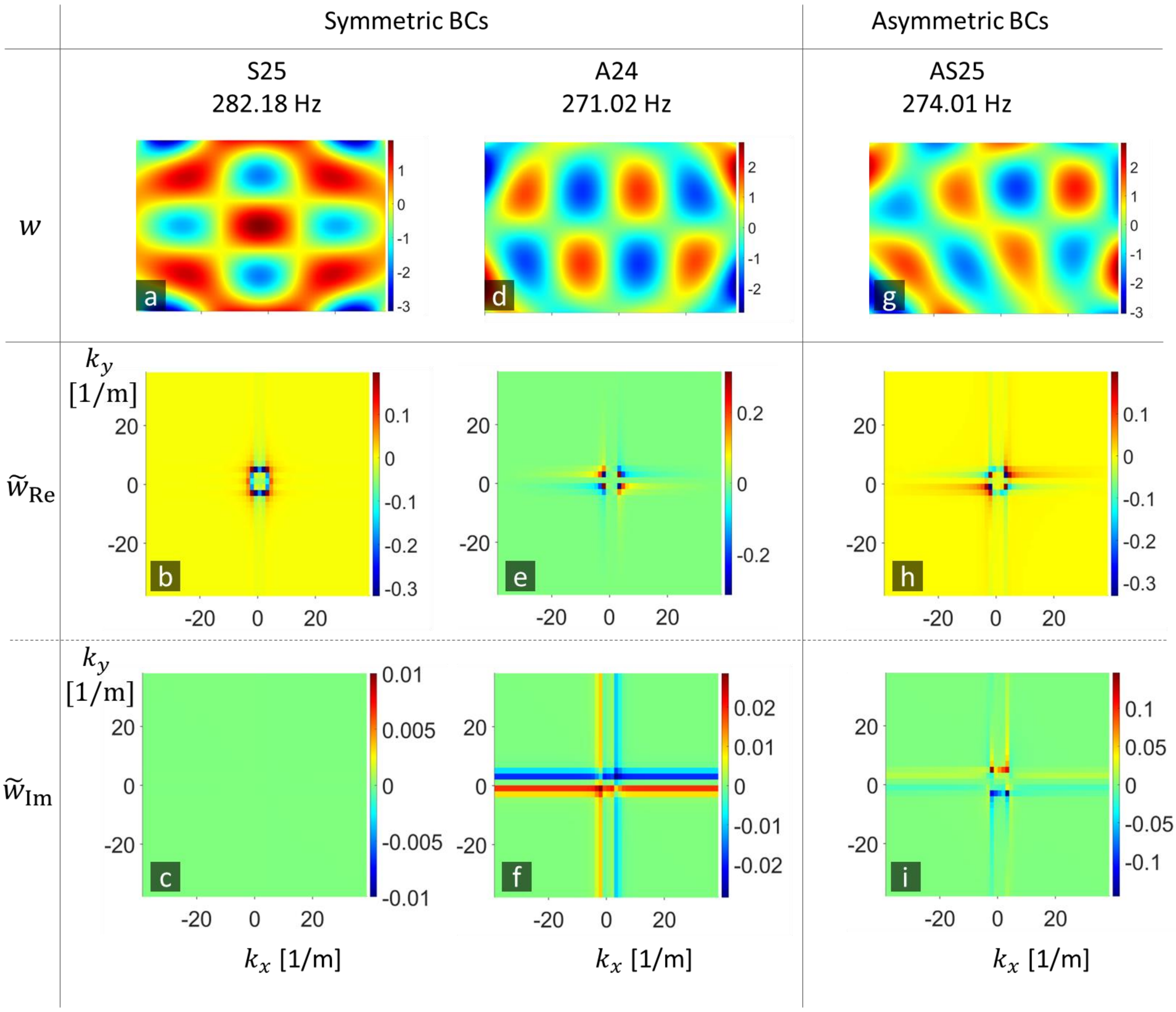

Figure 2. Selected mode shapes of the CFRP plate and corresponding $k$-space coefficient images computed by fast Fourier transform (FFT), including: (a-c) symmetric mode S25 and (d-f) anti-symmetric mode A24 of the plate with supported four corners, and (g-i) asymmetric mode AS25 of the plate with free bottom-right corner.

## 3. Methodology

### 3.1. PINN and the Proposed k-PINN extension at a Glance

PINNs utilize the power of deep NNs for mapping a set of input parameters into an output response function, and incorporate the physics knowledge and the underlying governing equations, to ensure that the predictions respect the laws of physics. As such, the loss function of PINNs is typically composed of two distinctive types of loss terms: a data-driven type (which incorporates the measurement data and known boundary values), and a physics-informed type (which incorporates the governing PDEs throughout the domain, herein the vibrating plate, and along its boundaries).

The governing physics equations may be applied in the strong form, or alternatively in the weak form, i.e., conservation of the total potential energy [42–44]. In the weak form, the PDEs are imposed in an integral (i.e. reduced order) form over the entire domain, which also intrinsically satisfies the free boundary conditions of the problem. Previous studies have demonstrated the tendency of this approach to converge to the fundamental modes of the solution and its deficiency in solving high-order eigen mode shapes of the response in a buckling problem [44]. In this study, the strong form of the governing PDE, in addition to the traction-free BCs, are applied together for the reconstruction of vibrational mode shapes.

The proposed k-PINN is an extension of PINNs, which instead of directly mapping the response function from its (spatial) input parameters, learns its coefficient maps in $k$-space and projects them to the response function in a deterministic way. A standard PINN and a simplified form of k-PINN are schematically illustrated in Figure 3 and their characteristic differences are outlined, for case study of mapping the vibrational mode shapes of a plate in $xy$-plane.

In a standard PINN architecture (Figure 3(a)), the spatial distribution of the vibrational response $w$ at a modal frequency $f$ can be directly mapped from the spatial domain $x, y$ (and additionally the corresponding frequency $f$ if learning multiple modes). The data-driven terms of loss function can be directly computed from the output of NN, but the physics-informed terms additionally require derivatives of the response with respect to the 2D spatial domain to construct the governing PDEs. Therefore, several additional backpropagations are sequentially performed to obtain the high-order derivatives of the PDEs.

In contrast, the proposed k-PINN (Figure 3(b)) learns the complex solution in the $k$-space (introduced in section 2.3) for the modal frequency $f$ given as input. The spatial solution and its derivatives of any order can then be analytically reconstructed at once through the linear superposition of all spectral components via inverse DFT. The NN learns the real and imaginary coefficients of the response in $k$-space, and it is the corresponding set of wavenumbers that construct the vibrational mode shapes in a deterministic way, without being affected by the spectral bias of the chosen NN structure.

In fact, reconstruction in $k$-space provides a rigorous and interpretable domain for adjusting the complexity and wavenumber composition of the response. This is particularly interesting for reconstruction of vibrational responses, given their sparsity in $k$-space (see Figure 2), which enables k-PINN to efficiently, and simultaneously, map multiple modal frequencies. Moreover, reducing the solution space to its $k$-space definition with a finite bandwidth, makes the training computationally more efficient. This is because the entire spatial response and its derivatives are analytically derived at once from the $k$-space solution, which is obtained from one forward solution of the NN. This means that the relevant loss terms can be readily and efficiently computed without requiring an additional set of sequential backpropagations for generating the derivative terms of the governing PDEs, and independent of the complexity of the NN.

As illustrated in Figure 3(b), an adaptive regularization mask is applied to the $k$-space solution obtained from the NN, which is meant for faster convergence to the desired solution space. Furthermore, to promote the sparsity and to obtain a compressed spectral mapping of the vibrational response, an additional sparsity loss term is considered in $k$-space. Detailed explanation of these steps and the relevant formulations are provided in the following section.

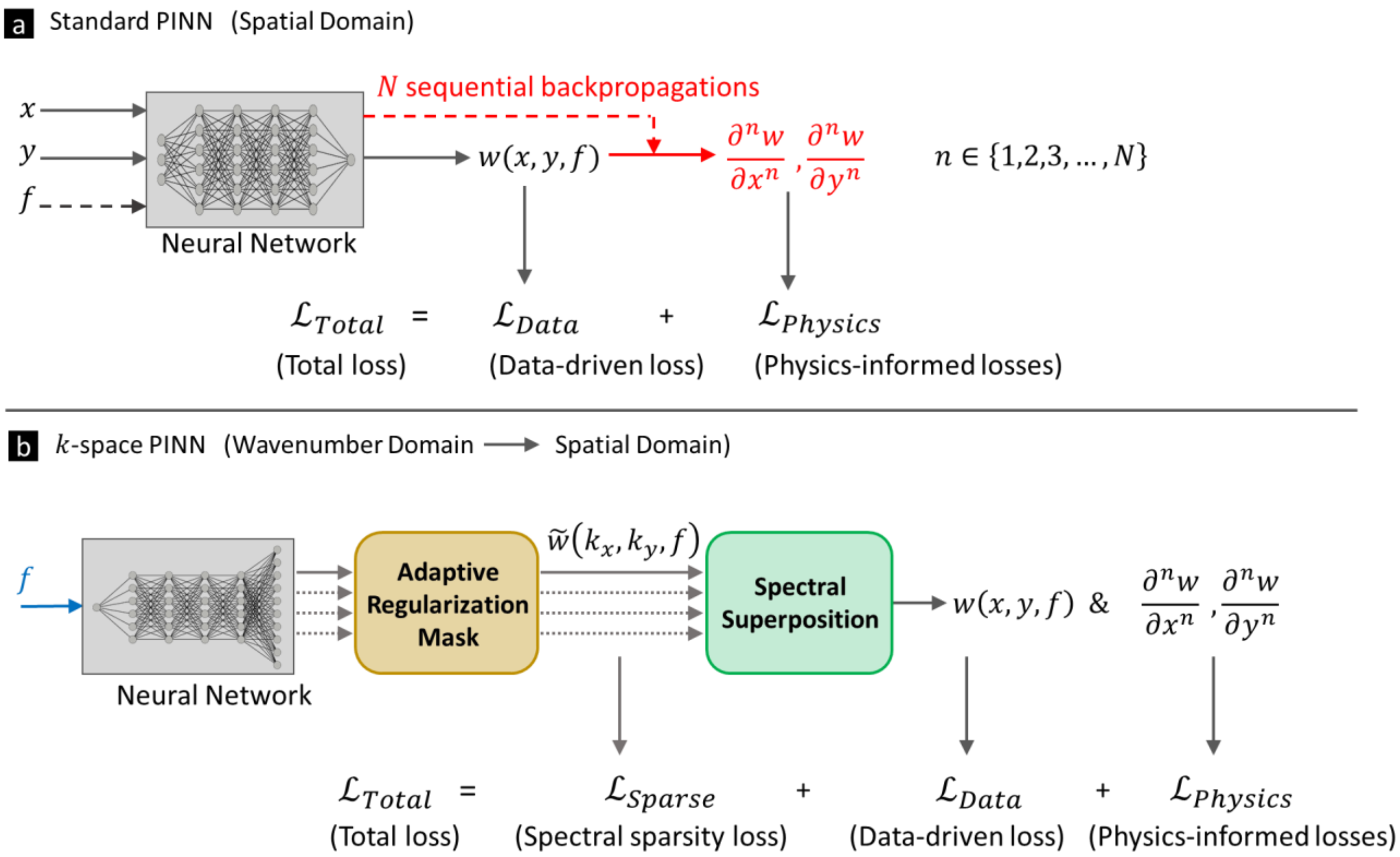

Figure 3. Schematic illustration of (a) a standard physics-informed neural network (PINN) in spatial domain, and (b) the proposed $k$-space PINN (k-PINN) in a simplified form, for mapping vibrational mode shapes of a plate in $xy$-plane at a frequency $f$

### 3.2. Detailed Architecture and Formulation of k-PINN

The flow chart of the proposed k-PINN methodology, along with its detailed architecture for addressing the bending vibrations of thin plates, is shown in Figure 4.

The first step is concerned with the collection of vibrational response from a set of (sparse) measurement points on the surface of the plate, and the identification of modal frequencies and relevant modal deflections (e.g. through FFT analysis and peak picking, or using PolyMAX method [45]). In practice, mode shapes may be naturally stimulated due to the presence of in-service loads, or externally stimulated using an excitation source, such as using an impact hammer. In the current study, this step is replaced with FE modal analysis of the plate and the addition of virtual noise to the simulation dataset, as explained in section 2.2.

Afterward, the modal frequencies and deflections are normalized and fed into k-PINN for full-field reconstruction of mode shapes and identification of unknown elastic properties through four subsequent stages, which will be introduced in the remainder of this section. Subsections relevant to each step are indicated in Figure 4, a brief outline of which is as follows: In section 3.2.1, the $k$-space definition of the vibrational response and its spatial derivatives, based on the normalized and centralized spatial domain are substantiated. In section 3.2.2, the governing $4^{th}$ order PDE is reparametrized and reduced to a $2^{nd}$ order PDE, by introducing $2^{nd}$ order derivatives of deflection (i.e. curvatures) as auxiliary output parameters. In sections 3.2.3, the NNs mapping of the response and auxiliary derivatives in $k$-space are formulated, and further in 3.2.4 their regularization using a mode-based adaptive Gaussian mask is introduced. Lastly, in section 3.2.5 the total loss function and optimization problem for simultaneous reconstruction of multiple mode shapes are formulated.

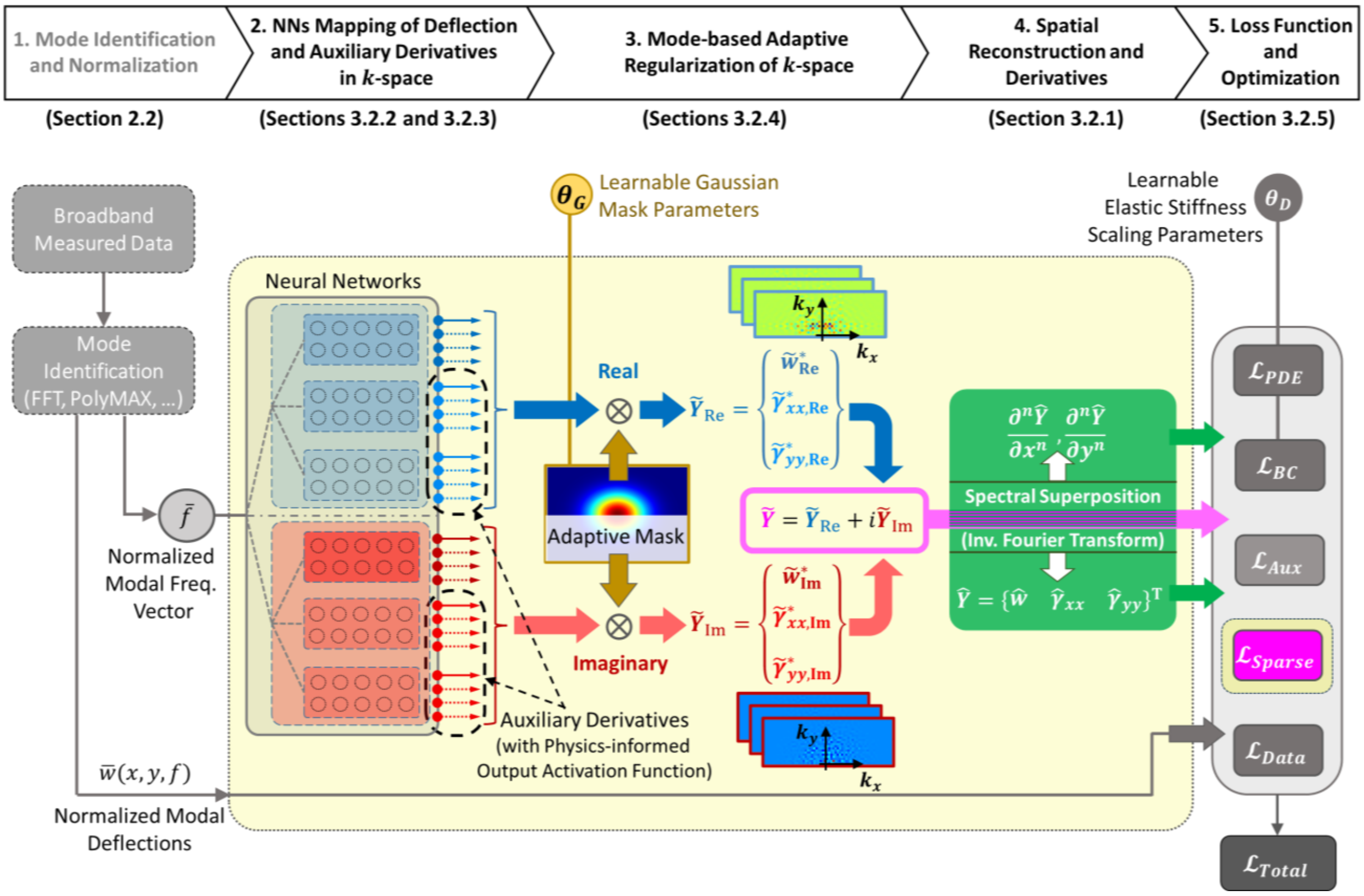

Figure 4. The architecture of proposed k-PINN for $k$-space mapping and reconstruction of multiple vibrational mode shapes. Multiple NNs can be considered for mapping real and imaginary components of the principal and the auxiliary output parameters, as illustrated with the dash lines inside the NNs box.

### *3.2.1. Spectral Formulation and Derivatives in $k$-space*

A spatial solution $\hat{w}(x, y, f)$ can be reconstructed from the corresponding complex form in $k$-space $\tilde{w}(k_x, k_y, f)$ by linear superposition of all individual spectral components through inverse DFT:

$$\hat{w}(x, y, f) = \sum_{m_x=-N_f\times M_x}^{N_f\times M_x} \sum_{m_y=-N_f\times M_y}^{N_f\times M_y} \tilde{w}\big(k_x(m_x), k_y(m_y), f\big) \exp\Big(2\pi i\big(k_x(m_x)x + k_y(m_y)y\big)\Big) \quad (14)$$

Subsequently, the spatial derivatives of any order $n$ can be directly derived as:

$$\begin{Bmatrix} \dfrac{\partial^n \hat{w}}{\partial x^n} \\ \dfrac{\partial^n \hat{w}}{\partial y^n} \end{Bmatrix} = (2\pi i)^n \sum_{m_x=-N_f\times M_x}^{N_f\times M_x} \sum_{m_y=-N_f\times M_y}^{N_f\times M_y} \begin{Bmatrix} k_x^n \\ k_y^n \end{Bmatrix} \tilde{w}\big(k_x(m_x), k_y(m_y), f\big) \exp\Big(2\pi i\big(k_x(m_x)x + k_y(m_y)y\big)\Big) \quad (15)$$

Considering the complex conjugate symmetry of $k$-space, the reconstruction can be reduced to the following real value equation for the upper space:

$$\hat{w}(\bar{x},\bar{y},f)=\sum_{m_x=-N_f\times M_x}^{N_f\times M_x}\sum_{m_y=0}^{N_f\times M_y}\{\tilde{w}^*_{\mathrm{Re}}\quad \tilde{w}^*_{\mathrm{Im}}\}\begin{Bmatrix}\cos\left(2\pi\left(k_x(m_x)\bar{x}\,+k_y(m_y)\bar{y}\right)\right)\\ \sin\left(2\pi\left(k_x(m_x)\bar{x}\,+k_y(m_y)\bar{y}\right)\right)\end{Bmatrix} \tag{16}$$

where the superscript (*) indicates the coefficients corresponding to this reduced $k$-space, and $\bar{x}$ and $\bar{y}$ are the centralized and normalized spatial coordinates defined as:

$$\begin{cases}\bar{x}=\dfrac{x}{L_x}-0.5\\ \bar{y}=\dfrac{y}{L_y}-0.5\end{cases} \tag{17}$$

Consequently, by learning the real (i.e. cosine) and imaginary (i.e. sine) coefficient matrices, the spatial response can be reconstructed.

As demonstrated in Figure 5, in the centralized coordinate system defined by Equation 17, the origin is shifted to the center of the spatial domain. Consequently, the real (cosine) spectral components are exclusively symmetric, and the imaginary (sine) spectral components are exclusively anti-symmetric for any integer or fractional wavenumber. This centralization enables the extraction of symmetric (anti-symmetric) components of the response using real-only (imaginary-only) coefficient matrices. Without this centralization, this exclusive symmetry condition of real and imaginary components is only valid at integer wavenumbers. Moreover, it can be observed how the addition of fractional mid-wavenumber components (i.e. $N_f=2$), enables the reconstruction of symmetric components with non-zero deflection at the edges (Figure 5(a)), and anti-symmetric components with zero deflection at the edges (Figure 5(b)).

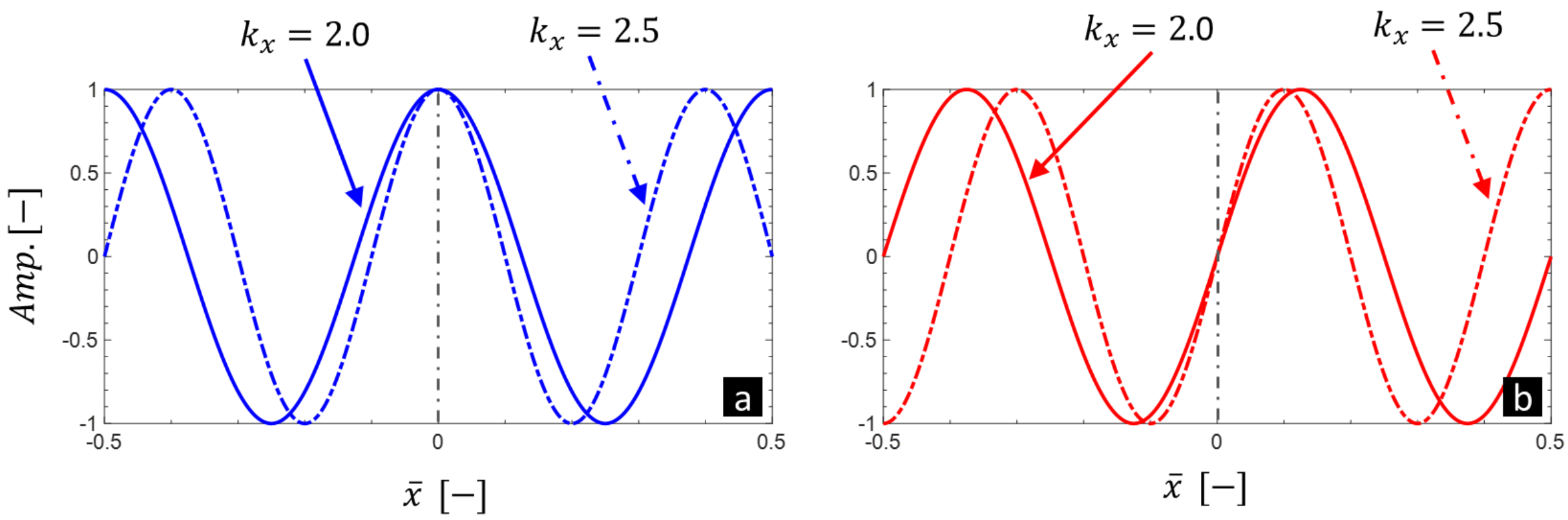

Figure 5. Examples of $k$-space components along the centralized and normalized spatial axis $\bar{x}$, with integer wavenumber $k_x=2.0$ and fractional mid-wavenumber $k_x=2.5$ (i.e. $N_f=2$), for (a) symmetric reconstruction with real (cosine) components, and (b) anti-symmetric reconstruction with imaginary (sine) components.

### 3.2.2. Auxiliary Derivatives and Reparametrized PDE

The governing equation of motion (Equation 7) is a 4th order PDE with a single degree of freedom, i.e., the deflection $w$. However, due to the noise amplification induced by the higher-order derivatives, the problem cannot converge to a valid solution in its current form, especially at higher frequencies. Therefore, the second-order derivatives (i.e., the curvatures $\gamma_{xx}$ and $\gamma_{yy}$ in Equation 2) are considered as auxiliary parameters, and the governing PDE is reformulated as a reduced 2nd order PDE as follows:

$$\mathcal{F} = -D_{11}\frac{\partial^2\gamma_{xx}}{\partial x^2} - 2(D_{12}+2D_{66})\frac{\partial^2\gamma_{xx}}{\partial y^2} - D_{22}\frac{\partial^2\gamma_{yy}}{\partial y^2} - I_d\omega^2 w + I_r\omega^2(\gamma_{xx}+\gamma_{yy}) = 0 \tag{18}$$

together with the two auxiliary 2nd order equations:

$$\begin{cases}\gamma_{xx} + \dfrac{\partial^2 w}{\partial x^2} = 0\\ \gamma_{yy} + \dfrac{\partial^2 w}{\partial y^2} = 0\end{cases} \tag{19}$$

So, the total response to be learnt in $k$-space, including the coefficient matrices of these auxiliary parameters, is as follows:

$$\widetilde{\boldsymbol{Y}} = \begin{Bmatrix}\widetilde{w}^*\\ \widetilde{\gamma}_{xx}^*\\ \widetilde{\gamma}_{yy}^*\end{Bmatrix} \tag{20}$$

This reparameterization, as proposed in a preceding study by [46], regularizes the high-order derivatives of the response function and has been shown to significantly improve the convergence of PDEs. This is particularly essential when inferring the elastic coefficients $D_{ij}$, as the corresponding terms are factorized by the highest (4th) derivative terms, which are more challenging to learn. Therefore, in the absence of such regularization, training most often converges to a trivial solution in which the governing equation $\mathcal{F}$ is minimized through fictitious near-zero elastic coefficients $D_{ij} \to 0$.

### 3.2.3. Neural Networks Mapping of $k$-space and Physics-informed Output Activation Function

In order to learn the mode shapes corresponding to a modal frequency, a fully connected deep neural network, as shown in Figure 4 is considered. This network takes frequency as the only input and has $6 \times (N_f \times M_x + 1) \times (N_f \times M_y + 1)$ outputs for mapping of the real and imaginary components of $\widetilde{\boldsymbol{Y}} = \{\widetilde{w}^* \quad \widetilde{\gamma}_{xx}^* \quad \widetilde{\gamma}_{yy}^*\}^{\mathrm{T}}$ corresponding to every individual wave vector $\boldsymbol{k} = \{k_x \quad k_y\}$. An alternative approach is a continuous mapping from $k$-space to the corresponding coefficient space, i.e., including $\boldsymbol{k} = \{k_x \quad k_y\}$ as input and mapping it to six outputs as the real and imaginary components of $\widetilde{\boldsymbol{Y}}$. It is noteworthy that we also implemented this latter approach in our research, but it was only the individual mapping of the components that was able to successfully construct $k$-space due to its discreteness and sparsity.

For the sake of learning efficiency, the input frequency vector $\boldsymbol{f}$ is centralized and normalized as follows:

$$\bar{\boldsymbol{f}} = \frac{\boldsymbol{f} - \min \boldsymbol{f}}{\max \boldsymbol{f} - \min \boldsymbol{f}} - 0.5 \tag{21}$$

Hence, considering a NN that includes $N$ hidden layers, the mapping from the normalized frequency vector $\bar{\boldsymbol{f}}$ at the first layer to the $k$-space response $\widetilde{\boldsymbol{Y}}_{\mathrm{NN}}$ in the last layer is as follows:

$$\boldsymbol{\ell_1} = g_H\left(a\boldsymbol{\beta_1}\left(\boldsymbol{\chi_1}\bar{\boldsymbol{f}} + \boldsymbol{b_1}\right)\right) \tag{22}$$

$$\widetilde{\boldsymbol{Y}}_{\mathrm{NN}} = g_O\left(\boldsymbol{\chi_{N+1}}.\boldsymbol{L_{N,xx}} + \boldsymbol{\ell_{N+1}}\right) \tag{23}$$

where $g_H$ is the activation function of hidden layers, with a constant scaling hyper-parameter $a$ and a neuron-wise adaptive scaling parameter $\boldsymbol{\beta_n}$ [32], and $\boldsymbol{\chi}_n$, $\boldsymbol{b}_n$ and $\boldsymbol{\ell_n}$ respectively are the weight, bias, and output parameters, and $n$ denotes the layer number. $g_O$ is the activation function of the output layer, which can be a linear function $g_O(x) = x$.

As explained in section 3.2.2, the auxiliary parameters $\tilde{\gamma}^*_{xx}$ and $\tilde{\gamma}^*_{yy}$ represent the curvatures, i.e., the second-order derivatives of bending deflection $\widetilde{w}^*$. According to Equation 15, these second-order derivatives are obtained in $k$-space through a linear superposition of the spectral components of deflection, each quadratically scaled by the corresponding squared circular wavenumber, i.e., factorized by $(2\pi k_x)^2$ for obtaining $\tilde{\gamma}^*_{xx}$ and factorized by $\left(2\pi k_y\right)^2$ for obtaining $\tilde{\gamma}^*_{yy}$. Therefore, to ensure efficient learning of the auxiliary parameters, proper scaling of relevant output coefficients during training is essential. This is particularly crucial for the higher frequency mode shapes, which have a more dominant contribution from higher wavenumbers, leading to a significantly higher magnitude of the second-order derivatives. While a straightforward approach might involve factorization the coefficients by their corresponding squared wavenumbers, this would lead to a predominant effect from higher wavenumbers and compromise the learning efficiency of lower frequency mode shapes.

To this end, a customized physics-informed output activation function $g_O$ is proposed for efficient reconstruction of broadband selection of mode shapes (section 4.3), which is a linear activation for mapping the coefficients of the bending deflection $\widetilde{w}^*$, and a nonlinear activation for mapping the auxiliary parameters $\tilde{\gamma}^*_{xx}$ and $\tilde{\gamma}^*_{yy}$:

$$g_O(a(i_k)) = \begin{cases} a(i_k) & , \quad a(i_k)\epsilon\Omega_{\widetilde{w}^*} \\ 2\pi k_x^{eff}(i_k)\, a(i_k) \exp\left(\left|\lambda_g a(i_k)\right|\right) & , a(i_k)\epsilon\Omega_{\tilde{\gamma}^*_{xx}} \\ 2\pi k_y^{eff}(i_k)\, a(i_k) \exp\left(\left|\lambda_g a(i_k)\right|\right) & , a(i_k)\epsilon\Omega_{\tilde{\gamma}^*_{yy}} \end{cases} \tag{24}$$

where $a(i_k)$ is the output element corresponding to the wavenumber component $i_k$, and $\lambda_g = 5$ is a hyperparameter chosen for a faster convergence, and:

$$k^{eff} = \begin{cases} 1 & , k = 0 \\ k & , k \neq 0 \end{cases} \quad (25)$$

The proposed output activation function enables an exponential amplification of auxiliary coefficients, and additionally localizes the factorization of each auxiliary coefficient with the relevant effective wavenumber. The effective wavenumber $k^{eff}$ is defined in such a way that the zero-wavenumber components are not removed by the customized output activation function, as defined in Equation 24.

### ***3.2.4. Mode-based Adaptive Regularization of $k$-space***

The reconstruction error of k-PINN (i.e. $\boldsymbol{Y} - \widehat{\boldsymbol{Y}}$ ) depends on the assumed bandwidth $\{B_x, B_y\}$ and the periodicity of the boundary values. Generally, the wider the bandwidth, the higher the reconstruction accuracy. Moreover, the better a mode shape can be periodized through the Fourier basis functions, the higher the reconstruction accuracy.

However, enlarging the bandwidth introduces many wavenumbers to the reconstructed response, which may introduce high spatial noise and slow down the convergence to the correct solution. In order to tackle this issue, the coefficient matrices $\widetilde{\boldsymbol{Y}}$ are regularized by factorizing the output matrices of the neural network $\widetilde{\boldsymbol{Y}}_{\mathrm{NN}}$ using a normalized bivariate Gaussian mask $\boldsymbol{G}$ as follows:

$$\widetilde{\boldsymbol{Y}} = \widetilde{\boldsymbol{Y}}_{\mathrm{NN}} G \quad (26)$$

$$G(k_x, k_y, f) = \frac{1}{\eta_n} \exp\left( -0.5 \left( \left( \frac{m_x}{s_{Gx}(f)} \right)^2 + \left( \frac{m_y}{s_{Gy}(f)} \right)^2 \right) \right) \quad (27)$$

where $s_{Gx}$ and $s_{Gy}$ are the frequency-dependent standard deviations of the Gaussian mask, which are adaptively changed during the learning process for each individual mode shape, and $\eta_n$ is a normalization factor that ensures the mask applies a maximum scaling of unity.

The standard deviations are mapped from two learnable parameters $\{\theta_{Gx}, \theta_{Gy}\}$, such that they are constrained within a given radius $\{r_{Gx}, r_{Gy}\}$ relative to given initial values $\{s_{Gx0}, s_{Gy0}\}$:

$$\begin{Bmatrix} s_{Gx} \\ s_{Gy} \end{Bmatrix} = \begin{Bmatrix} (s_{Gx0} - r_{Gx}) + \dfrac{2r_{Gx}}{1 + exp\,(-\alpha\theta_{Gx})} \\ (s_{Gy0} - r_{Gy}) + \dfrac{2r_{Gy}}{1 + exp\,(-\alpha\theta_{Gy})} \end{Bmatrix} \quad (28)$$

where $\alpha$ is a scaling factor that determines the variation rate of $\{s_{Gx}, s_{Gy}\}$ in relation to the learnable parameters $\{\theta_{Gx}, \theta_{Gy}\}$. A scaling factor of $\alpha = 3$ ensures that the standard deviation can reach a close proximity of its lower and upper bounds within a parameter range of $[-1,1]$. The constants $\{s_{Gx0}, s_{Gy0}\}$ and $\{r_{Gx}, r_{Gy}\}$ have to be chosen such that the Gaussian mask initially narrows down the effective bandwidth of $k$-space, while it can be expanded to adequately cover the entire bandwidth when reaching the upper bound of its standard deviation. As an example, setting $s_{Gx0} = 4$ and $r_{Gx} = 2$, leads to an upper bound of $s_{Gx0} + r_{Gx} = 6$ which means a bandwidth of $B_x = 18$ can be covered within a three-sigma range of the Gaussian mask. The mapping of learnable parameter $\theta_{Gx}$ to the standard deviation $s_{Gx}$ for these hyperparameter settings (as applied in this study) is demonstrated in Figure 6.

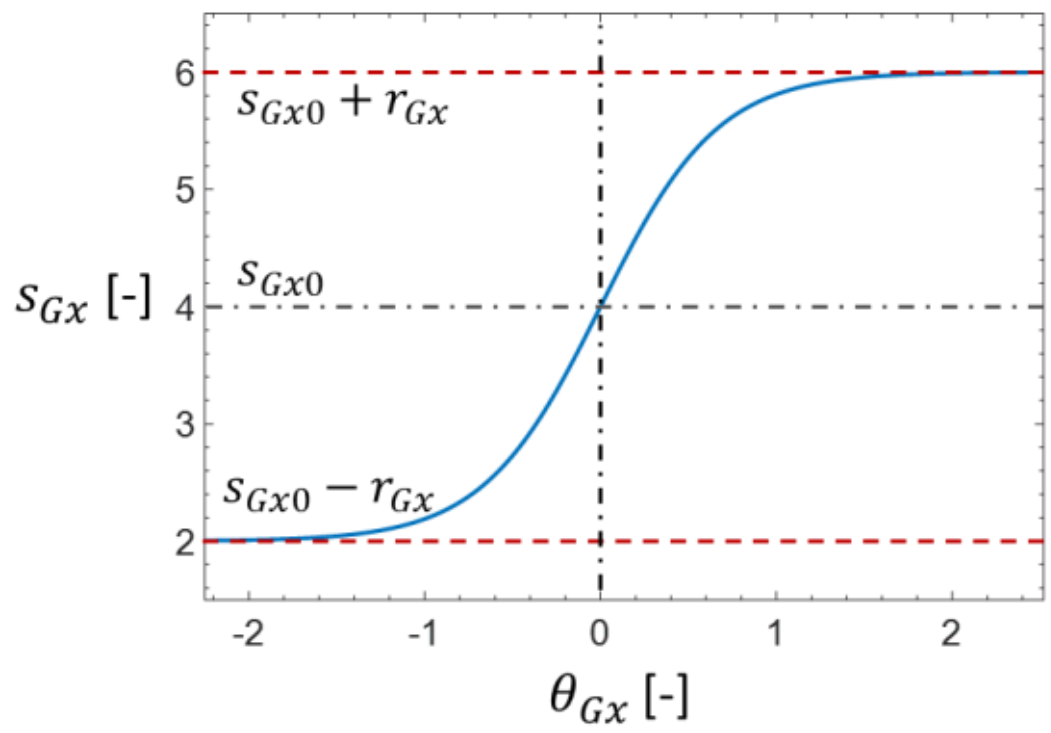

Figure 6. Mapping of the learnable parameter $\theta_{Gx}$ to the standard deviation $s_{Gx}$ for a scaling factor of $\alpha = 3$, and an initial value $s_{G0} = 4$ and bounding radius of $r_{Gx} = 2$, as used in this study.

The applied zero-mean Gaussian mask is, in fact, an adaptive low-pass filter applied in $k$-space, which is considered for the reconstruction of relatively low-frequency vibrational mode shapes in this case study. One may consider other definitions when dealing with other problems, e.g. considering a Gaussian ring, for bandpass filtering over a particular wavenumber bandwidth for the reconstruction of narrowband, high-frequency vibrations.

#### *3.2.5. Loss Functions*

After obtaining a regularized $k$-space response and substituting it into Equation 16, the spatial response and its derivatives are computed and plugged into a loss function $\mathcal{L}_{Total}$ to be minimized:

$$\underset{\boldsymbol{\beta},\boldsymbol{\chi},\boldsymbol{b},\boldsymbol{\theta_G},\boldsymbol{\theta_D}}{\operatorname{argmin}} \mathcal{L}_{Total} \tag{29}$$

where $\boldsymbol{\beta}, \boldsymbol{\chi}$ and $\boldsymbol{b}$ are the inherent scaling, weight, and bias parameters of the NN, $\boldsymbol{\theta_G} = \{\theta_{Gx} \quad \theta_{Gy}\}$ represents the vector of additional learnable parameters corresponding to the Gaussian mask; and $\boldsymbol{\theta_D} = \{\theta_{D11} \quad \theta_{D22} \quad \theta_{D12} \quad \theta_{D66}\}$ denotes the vector of additional learnable scaling factors of the elastic coefficients $D_{ij}$ relative to a set of pre-trained/initial-guess values $D_{ij}^0$ such that:

$$D_{ij} = \theta_{Dij} D_{ij}^0 \tag{30}$$

To construct the loss function $\mathcal{L}_{Total}$, a set of data-driven, physics-informed, auxiliary, and sparsity-promoting loss terms are considered as follows:

$$\begin{cases} \mathcal{L}_{\mathrm{Data}} = \|\widehat{w} - \overline{w}_T\|_{\Omega_{\mathrm{Data}}} \\ \mathcal{L}_{\mathrm{PDE}} = \|\mathcal{F}\|_{\Omega_{\mathrm{CP}}} \\ \mathcal{L}_{\mathrm{BC}} = \left\|\widehat{M}_{nn}\right\|_{\Omega_{\mathrm{B}}} + \left\|\widehat{V}_n\right\|_{\Omega_{\mathrm{B}}} + \|\widehat{w}\|_{\Omega_{\mathrm{B}}} \\ \mathcal{L}_{\mathrm{Sparse}} = |\widetilde{w}^*|_{\Omega_{\mathrm{k}}} \\ \mathcal{L}_{\mathrm{Aux}} = \begin{cases} \left\|\widehat{\gamma}_{xx} + \dfrac{\partial^2 \widehat{w}}{\partial x^2}\right\|_{\Omega_{\mathrm{CP}}} + \left\|\widehat{\gamma}_{yy} + \dfrac{\partial^2 \widehat{w}}{\partial y^2}\right\|_{\Omega_{\mathrm{CP}}} & \text{(spatial domain)} \\ & \text{OR} \\ \|\widetilde{\gamma}_{xx}^* + k_x^2 \widetilde{w}^*\|_{\Omega_k} + \left\|\widetilde{\gamma}_{yy}^* + k_y^2 \widetilde{w}^*\right\|_{\Omega_{\mathrm{k}}} & (k-\text{space}) \end{cases} \end{cases} \tag{31}$$

where $\|.\|$ denotes the mean square error (MSE) defined as $\|g\|_{\Omega} = \frac{1}{N_{\Omega}} \sum_{n=1}^{N_{\Omega}} g(x_n, y_n, f_n)^2$, and $|.|$ denotes the $l1$ norm in the complex $k$-space defined as $|g|_{\Omega} = \frac{1}{N_{\Omega}} \sum_{n=1}^{N_{\Omega}} \sqrt{g_{Re}^2 + g_{Im}^2}$, for $N_{\Omega}$ evaluation points over the domain $\Omega$. The symbol $(\hat{})$ denotes the reconstructed response obtained from k-PINN, and $\overline{w}$ denotes the measured response of modes normalized with respect to their maximum, such that for every measurement point $n$:

$$\overline{w}_T(x(n), y(n), f) = \frac{w_T(x(n), y(n), f)}{\max_j w_T(x(j), y(j), f)} \tag{32}$$

$\mathcal{L}_{\mathrm{Data}}$ is the data-driven loss term enforcing the measurement data over the domain $\Omega_{\mathrm{Data}}$, $\mathcal{L}_{\mathrm{PDE}}$ is the physics-informed loss term enforcing the governing PDE (Equation 18) over a set of collocation points (CPs) in the domain $\Omega_{\mathrm{CP}}$, $\mathcal{L}_{\mathrm{BC}}$ is the physics-informed loss term enforcing the traction-free boundary conditions and the supported corners over the boundary domain $\Omega_{\mathrm{B}}$. $\mathcal{L}_{\mathrm{Sparse}}$ is a loss term applied in $k$-space, which aims to maximize the sparsity of $k$-space by minimizing the $l1$ norm of its coefficients, so that a compressed spectral representation of the solution is achieved.

$\mathcal{L}_{\mathrm{Aux}}$ is an auxiliary loss term required to ensure the validity of the reparametrized PDE introduced in section 3.2.2. $\mathcal{L}_{\mathrm{Aux}}$ can be applied in the spatial domain $\Omega_{\mathrm{CP}}$, by imposing the set of auxiliary Equations 19 (i.e., $\gamma_{xx} + \frac{\partial^2 w}{\partial x^2} = 0$ and $\gamma_{yy} + \frac{\partial^2 w}{\partial y^2} = 0$). Given the spectral derivative Equation 15, $\mathcal{L}_{\mathrm{Aux}}$ can also be applied in $k$-space $\Omega_{\mathrm{k}}$ (i.e., by imposing $\widetilde{\gamma}_{xx}^* + k_x^2 \widetilde{w}^* = 0$ and $\widetilde{\gamma}_{yy}^* + k_y^2 \widetilde{w}^* = 0$ for all spectral components). The effect of these two formulations on the performance of k-PINN is studied in section 4.3.

The five different types of loss terms (data-driven and physics-informed) introduced in Equation 31, and the loss terms of the same type obtained for different modes, all can have significantly different magnitudes. Generally, a higher modal frequency is associated with higher wavenumbers, and consequently, a higher spatial gradient of PDE terms. Moreover, the inertial terms of the governing PDE (Equation 18) scale quadratically with frequency. Therefore, proper balancing of all these loss terms is essential for efficient training of k-PINN. To this end, each loss type for every individual mode is initially normalized to its magnitude at the start of training $(epoch = 0)$, and subsequently, the total loss function of k-PINN is defined as the mean of normalized losses:

$$\mathcal{L}_{Total} = \frac{1}{5}\left(\bar{\mathcal{L}}_{Data} + \bar{\mathcal{L}}_{PDE} + \bar{\mathcal{L}}_{BC} + \bar{\mathcal{L}}_{Aux} + \bar{\mathcal{L}}_{Sparse}\right) \tag{33}$$

where the symbol $(\bar{\ })$ denotes the normalized loss terms. Hence, the total loss is always initialized to unity, with equal weights of the individual loss terms. Various approaches have been proposed for efficient and adaptive balancing of the loss terms in PINN, for example [30; 47], and their performance for k-PINN could be further investigated.

One may convert the measured spatial data to $k$-space using 2D-FFT and apply the data-driven loss term $\mathcal{L}_{\mathrm{Data}}$ in $k$-space, provided that the data is non-sparse and has sufficient spatial resolution.

## 4. Results and Discussion

In this section, the performance of k-PINN in learning and inversion of the vibrational problem introduced in section 2 is evaluated. Initially, the performance of k-PINN, in comparison to standard (i.e., spatial domain) PINN for the reconstruction of a single mode shape, is compared. Then, k-PINN is applied to a set of case studies, considering the effect of sparsity and noisiness of measurement data, and their performance in the inversion of elastic coefficients from multiple modal frequencies is demonstrated.

Addition of noise to the test dataset makes a direct comparison of the true simulated response $w_{\mathrm{FEA}}$ and the reconstructed response $\widehat{w}$ irrelevant. Hence, to evaluate the reconstruction error of mode shapes, two standardized measures are introduced as follows:

$$\mathcal{E}_{\widehat{w}}(m) = 1 - \left|\sum_{i=1}^{N_s} w_{\mathrm{FEA}}^{\mathrm{STD}}(i,m)\, \widehat{w}^{\mathrm{STD}}(i,m)\right| \tag{34}$$

$$\mathcal{E}_{\widehat{w}_i}(i,m) = \frac{\left|w_{\mathrm{FEA_Thin}}^{\mathrm{STD}}(i,m) - \widehat{w}^{\mathrm{STD}}(i,m)\right|}{\max_i \left|w_{\mathrm{FEA}}^{\mathrm{STD}}(i,m)\right|} \tag{35}$$

where $g^{\mathrm{STD}}(i,m) = \frac{g(i,m)}{\sqrt{\sum_{j=1}^{N_s} g^2(j,m)}}$ denotes the standardized form of mode shapes, $\mathcal{E}_{\widehat{w}}(m)$ is the overall miscorrelation of reconstructed mode $m$, and $\mathcal{E}_{\widehat{w}_i}(i,m)$ is the relative reconstruction error at every individual spatial point $i$ of it.

**4.1. Architecture and Hyperparameters**

Fully connected NNs with locally adaptive sinusoidal activation functions are used throughout this study. A constant activation scaling factor of $a = 1$ is used for k-PINN, as it is the deterministic $k$-space bandwidth that determines the spatial frequency of reconstruction, not its activation scaling. However, different activation scaling factors of 1, 2 and 5 are examined for standard PINN to manipulate their output spatial frequency. In any case, the effective scaling factor allows for fine-tuning of this scaling at every individual neuron during the training. Different NN widths $(l)$ of 32, 128 and 512 are evaluated, with a constant number of five hidden layers.

By default, a bandwidth of $\{B_x, B_y\} = \{15,10\}$ is applied in $k$-space (for the rectangular plate of dimensions $750 \times 500\ \mathrm{mm}^2$), and fractional mid-wavenumber components are considered by setting $N_f = 2$. This sums up to 1281 spectral components in the upper independent half of $k$-space. Considering this bandwidth, wavelengths as small as $\frac{750}{15} = \frac{500}{10} = 50\ \mathrm{mm}$ can be accurately reconstructed. The Gaussian mask is initialized with standard deviations $\{s_{Gx0}, s_{Gy0}\} = \{4.0,4.0\}$ and bounding radii of $\{r_{Gx}, r_{Gy}\} = \{2.0,2.0\}$.

Both uniform and random sampling strategies of collocation points (CPs) are examined. By default, a uniform grid of $45 \times 30$ CPs is considered over the plate's surface in the $xy$-plane so that the spectral component with the shortest wavelength (i.e., the highest wavenumber) can be sampled with more than two points. In the case of random sampling, the same number of CPs $(45 \times 30 = 1350)$ are resampled at each training epoch using Latin hypercube method. In theory, random incoherent resampling of CPs during a sufficiently large number of training epochs should provide enough evaluation points for learning the spectral components of any wavelength.

The reparameterization of the governing PDE and related auxiliary parameters (introduced in section 3.2.2) are equally considered for both PINN and k-PINN. By default, the auxiliary loss function is calculated over CPs in the spatial domain, i.e., $\Omega_{\mathrm{CP}}$ as defined in Equation 31.

The training is performed using ADAM optimizer with a given number of epochs using the full batch of CPs, and a constant learning rate of $2 \times 10^{-4}$ is considered. The implementation is done in Python using the Pytorch deep learning library.

### 4.2. PINN versus k-PINN

The symmetric vibrational mode S15, with a relatively low modal frequency of 150.01 Hz, is selected for evaluating the performance of PINN and k-PINN. The mode shape, with the addition white Gaussian noise (SNR=20 dB), is shown in Figure 7(a). We start the evaluation with a simplified problem, in which it is assumed that only a single overall stiffness scaling factor, $\bar{\theta}_D$, is unknown, i.e., $\{\theta_{D11} \quad \theta_{D22} \quad \theta_{D12} \quad \theta_{D66}\} = \{\bar{\theta}_D \quad \bar{\theta}_D \quad \bar{\theta}_D \quad \bar{\theta}_D\}$. The training is initiated with an initial guess of $\bar{\theta}_D = 0.5$ (which must converge to the correct value of $\bar{\theta}_D = 1.0$), and it is continued for $50 \times 10^3$ epochs.

Table 2 summarizes the results obtained from all scenarios, and Figure 7 compares the test dataset with the mode shapes reconstructed by the with $l = 128$. A uniform grid of $12 \times 8 = 96$ measurement points is used for training as indicated with circular markers. No data points are considered along the free edges, as they may not be easily accessible in practice. This also evaluates the efficiency of boundary value reconstruction in the absence of data.

Table 2. Performance of PINN and k-PINN for vibrational mode 15, with various hyperparameter settings and a constant number of five hidden layers, after $50 \times 10^3$ training epochs

| NNs' Type | $l$ | $a$ | $\log_{10} \varepsilon_{\hat{w}}$ | $\bar{\theta}_D$ |
|---|---|---|---|---|
| PINN | 32 | 1 | -0.037 | 0.426 |
| | | 2 | -0.098 | 0.375 |
| | | 5 | -0.062 | 0.568 |
| | 128 | 1 | -0.040 | 0.437 |
| | | 2 | -0.087 | 0.385 |
| | | 5 | -0.049 | 0.451 |
| | 512 | 1 | -0.035 | 0.451 |
| | | 2 | -0.109 | 0.387 |
| | | 5 | -0.101 | 0.441 |
| k-PINN | 32 | 1 | -2.752 | 0.996 |
| | 128 | 1 | -2.831 | 0.995 |
| | 512 | 1 | -2.870 | 0.992 |

From the results, it is evident that only the k-PINN implementation can reconstruct Mode S15, and infer the unknown stiffness parameter $\bar{\theta}_D$ with relatively high accuracy, close to the true value of unity. The standard PINN fails to converge to a valid solution in all considered scenarios, consistently converging towards a lower stiffness scaling. The stiffness coefficients factorize the highest (4$^{th}$) order terms of the governing PDE (Equation 7), which are generally harder to regularize and learn. As such, PINN may trivially lower the stiffness to balance the elastic (high order) and inertial (low order) terms of PDE and minimize the physical loss.

Figure 7 clearly shows that PINN results in erroneous and totally uncorrelated spatial reconstruction of the mode shape, while k-PINN succeeds with a relatively small error. The reconstruction error of k-PINN (Figure 7(g)) is higher at the borders and is the highest in the vicinity of supported corners. This, in fact, represents an optimized response projected onto the reduced solution space prescribed by the bandwidth and resolution of $k$-space.

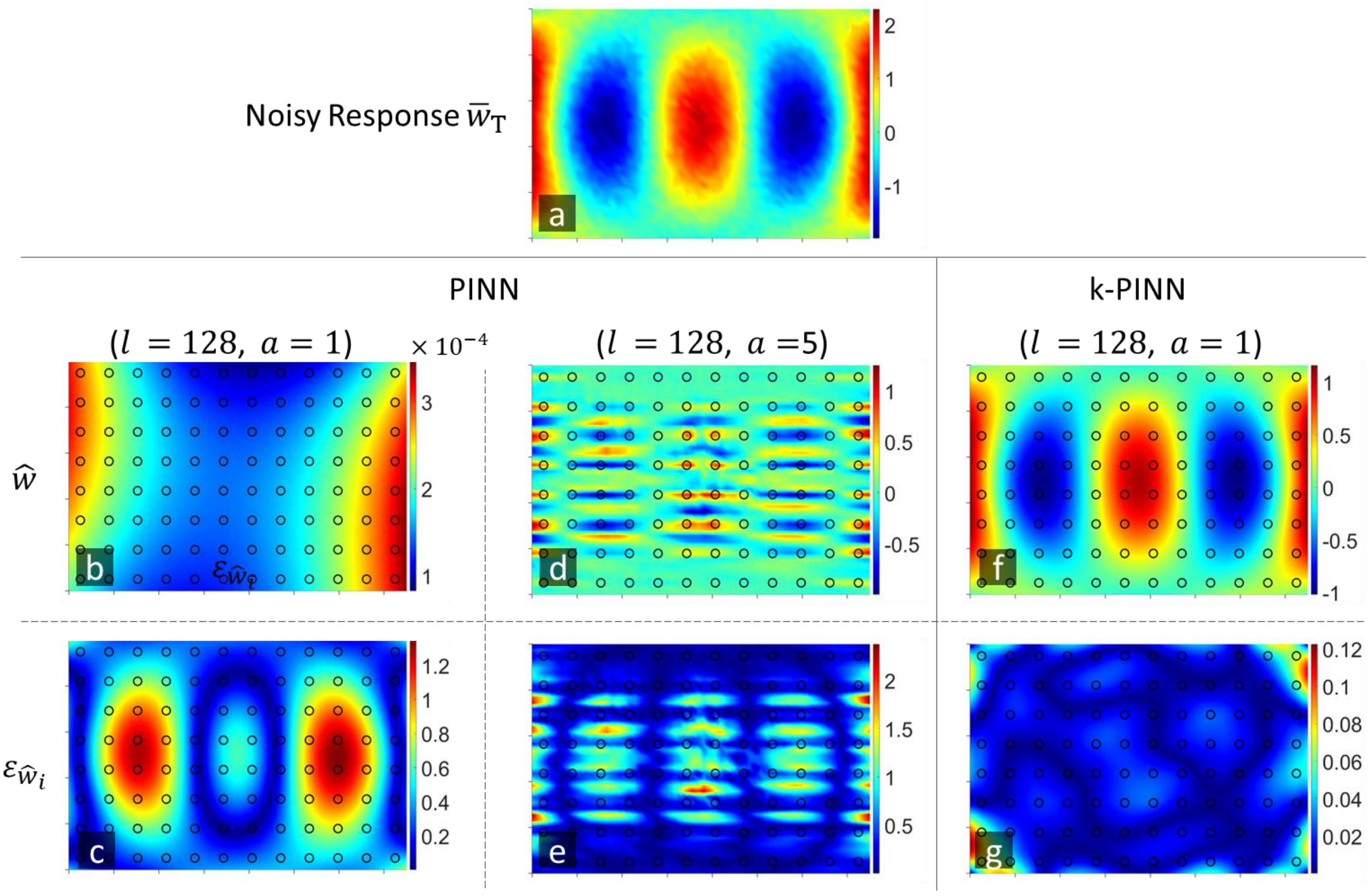

Figure 7. (a) Virtual noisy test data corresponding to the symmetric vibrational mode S15 (150.01 Hz), and its reconstruction with (b-e) PINN and (f,g) k-PINN, using a uniform grid of $12 \times 8 = 96$ measurement points indicated with circular markers. The middle row and the bottom row, respectively, show the reconstructed mode shape and corresponding error.

Furthermore, Figure 8 shows the evolution of the overall reconstruction error $\mathcal{E}_{\hat{w}}$ and the stiffness scaling factor $\bar{\theta}_D$ during training of k-PINN. The widest width of 512 leads to a lower reconstruction error but a slower convergence of the stiffness scaling, whereas a width of 128 results in a comparable reconstruction error but the fastest convergence of stiffness scaling, occurring at around 15’000 epochs.

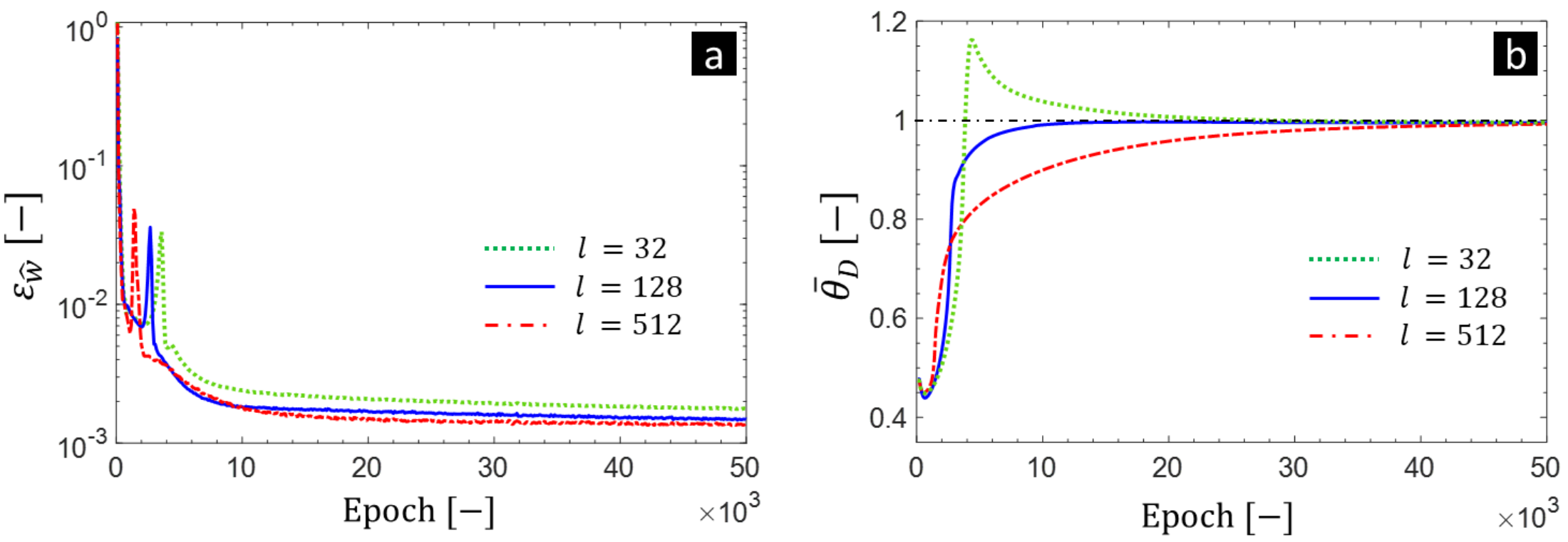

Figure 8. Performance of k-PINN for vibrational mode S15: (a) the reconstruction error $\mathcal{E}_{\hat{w}}$ and (b) the inferred overall stiffness scaling factor $\bar{\theta}_D$ from an initial guess of $\bar{\theta}_D = 0.5$, in function of training epochs

It must be noted that the purpose of this section is not to conclude that standard spatial PINN can never accurately learn such a vibrational mode shape. One may adjust the NN settings to tailor PINN to this particular problem or allow it to run for a larger number of training epochs, so that it might eventually converge to a valid solution. However, the results clearly demonstrate that, for the same problem formulation and NN structure, k-PINN accurately infers the unknown stiffness and reconstruct the full-field response, while the standard PINN fails across the range of settings examined herein. In the following sections, it is shown that k-PINN can simultaneously reconstruct a selection of multiple modal frequencies.

### 4.3. K-PINN for a Broadband Selection of Mode Shapes

#### *4.3.1.Symmetric BCs*

The focus of this section is on evaluating  the performance of k-PINN in simultaneously learning a relatively broadband selection of 10 mode shapes, as shown in Figure 9, obtained up to 1000 Hz for the plate with the symmetric BCs. The selection includes three mode shapes with rectangular symmetry and seven mode shapes with rectangular anti-symmetry.

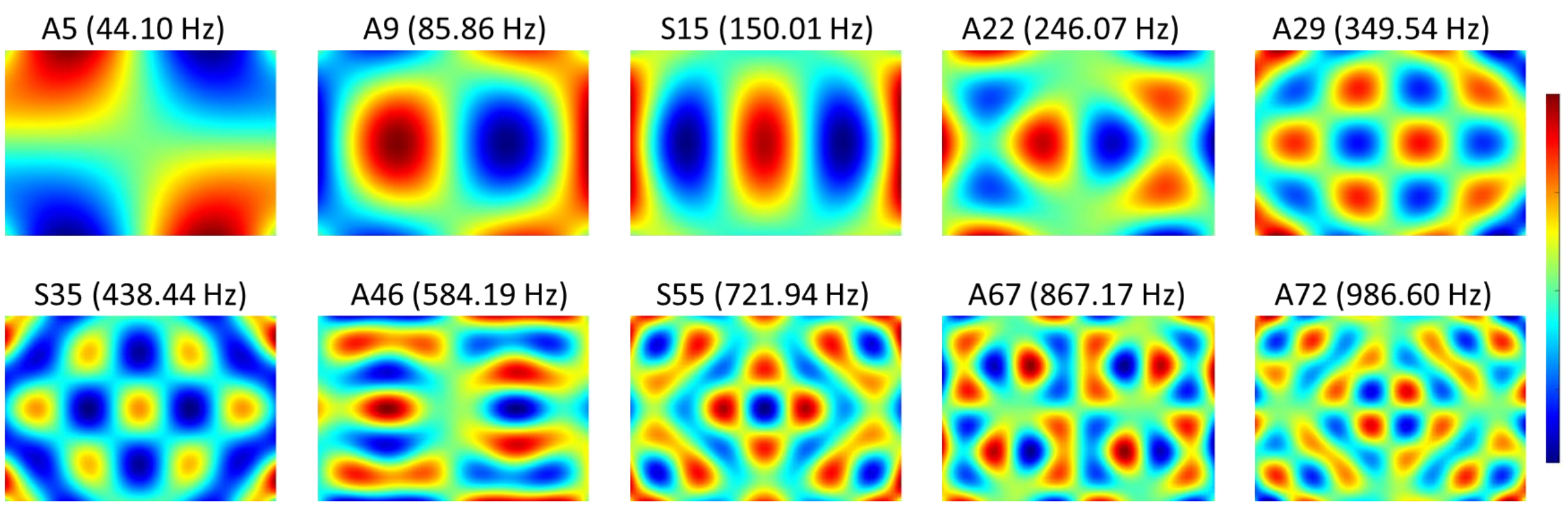

Figure 9. Selected broadband set of 10 modes for the plate with symmetric BCs, i.e. supported at the four corners.

An ablation study is conducted to demonstrate the contribution of different components of the proposed k-PINN methodology, including: the customized output activation function (section 3.2.3), the adaptive Gaussian mask (section 3.2.4), the spatial or $k$-space formulation of the auxiliary loss $\mathcal{L}_{Aux}$ and the addition of the sparsity loss $\mathcal{L}_{Sparse}$ (section 3.2.5). Moreover, different hyperparameter settings and distributions of data points are examined to understand their impact on the performance of k-PINN.

A constant NN width of $l = 128$ and depth of 5 hidden layers are maintained. Additionally, two NNs of the same architecture are used to independently map the real and the imaginary coefficients of $k$-space. For simplicity, and similar to the pervious section, a single overall stiffness scaling factor $\bar{\theta}_D$ is assumed as the only unknown material property to be identified from an initial underestimated guess of 0.5. Besides inferring the unknown stiffness scaling $\bar{\theta}_D$, the focus is on the performance of k-PINN in reconstructing the broadband set of mode shapes using different spatial distributions of test data. For this purpose, four test datasets are considered: a dense uniform grid of 45x30 data points (U45x30), a uniform grid of 96=12x8 data points (U12x8), and two randomly located sets of 96 (R96) and 48 (R48) data points. Cases with the same NN architecture are examined with a fixed initialization, for a fair comparison of their relative performance.

The results of the 14 examined case studies, denoted by CS1 to CS14, are summarized in Table 3. This summary includes the reconstruction error of three selected modes - S15, A46 and A72 - ranging from a relatively low to the highest modal frequency (see Figure 9). Moreover, Figure 10 shows the evolution of the reconstruction error for all individual modes and the evolution of the stiffness scaling $\bar{\theta}_D$ versus training epochs, for selected cases. The data-driven cases CS1&2 (i.e., without considering physics-informed loss terms), are trained for $50 \times 10^3$ epochs. The remaining physics-informed cases, CS3 to CS14, exhibit slower convergence and thus are trained for a larger number of $150 \times 10^3$ epochs. This slower convergence can be attributed to the additional constraints imposed by the physics-informed loss terms and the activation of auxiliary output parameters, which are , in fact, idle in the data-driven cases.

The first two cases, CS1&2, are dedicated to data-driven training, using a noise-free U45x30 dense dataset, both with and without the application of the Gaussian mask, using a single NN. According to the results (Figure 10(a,b)), in both scenarios, all mode shapes achieve a good reconstruction accuracy up until the training epoch 50,000. However, the application of the Gaussian mask (Figure 10(a)) significantly accelerates the convergence rate of all modes, while slightly increasing the reconstruction error due to the reduced solution space initialized by the mask.

Subsequently, physics-informed cases CS3 to CS6 are listed in Table 3, using the dense dataset U45x30 and with the addition of noise. From the results, it can be understood that after incorporating the physics-informed loss terms, a single NN ($N_{\mathrm{NN}} = 1$) is not as efficient (CS3), and considering two NNs ($N_{\mathrm{NN}} = 2$) improves the reconstruction accuracy of mode shapes (CS4), particularly at higher frequencies. Moreover, the results confirm that the addition of noise (CS5) barely affects the performance of k-PINN compared to the noise-free case CS4. Similar to the data-driven cases CS1&2, eliminating the Gaussian mask in the physics-informed case CS6 leads to delayed convergence of k-PINN and inaccurate stiffness scaling, also shown in Figure 10(c). Nonetheless, the reconstruction error achieved in the absence of the Gaussian mask is considerably lower (Figure 10(d)), which suggests a predominantly data-driven convergence of reconstruction given the relatively high density of data points.

Table 3. Performance of k-PINN for simultaneously learning the broadband selection of 10 modes with symmetric BCs (Figure 9). All NNs have five hidden layers and a constant width of 128.

| Case ID | NN settings | | Data | Noise | CPs | k-PINN settings | | | | $\log_{10} \varepsilon_{\hat{w}}$ | | | $\bar{\theta}_D$ |
|---|---|---|---|---|---|---|---|---|---|---|---|---|---|
| | $N_{\mathrm{NN}}$ | $g_O$ | | | | Phys. Loss | Gauss. Mask | Spars. Loss | Aux. Loss | S15 | A46 | A72 | |
| CS1 | 1 | NL | U60x40 | - | U | - | ✓ | ✓ | $\Omega_{\mathrm{CP}}$ | -2.84 | -2.58 | -2.51 | - |
| CS2 | 1 | NL | U60x40 | - | U | - | - | ✓ | $\Omega_{\mathrm{CP}}$ | -3.27 | -2.99 | -2.76 | - |
| CS3 | 1 | NL | U60x40 | - | U | ✓ | ✓ | ✓ | $\Omega_{\mathrm{CP}}$ | -3.04 | -2.79 | -2.16 | 0.993 |
| CS4 | 2 | NL | U60x40 | - | U | ✓ | ✓ | ✓ | $\Omega_{\mathrm{CP}}$ | -3.06 | -2.85 | -2.42 | 0.989 |
| CS5 | 2 | NL | U60x40 | ✓ | U | ✓ | ✓ | ✓ | $\Omega_{\mathrm{CP}}$ | -3.06 | -2.84 | -2.44 | 0.988 |
| CS6 | 2 | NL | U60x40 | ✓ | U | ✓ | - | ✓ | $\Omega_{\mathrm{CP}}$ | -3.05 | -2.80 | -2.51 | 0.761 |
| CS7 | 2 | NL | U12x8 | ✓ | U | ✓ | ✓ | ✓ | $\Omega_{\mathrm{CP}}$ | -2.81 | -2.51 | -1.86 | 0.980 |
| CS8 | 2 | NL | R96 | ✓ | U | ✓ | ✓ | ✓ | $\Omega_{\mathrm{CP}}$ | -2.53 | -2.44 | -1.74 | 0.989 |
| CS9 | 2 | NL | R96 | ✓ | U | ✓ | - | ✓ | $\Omega_{\mathrm{CP}}$ | -0.58 | -1.53 | -1.49 | 0.747 |
| CS10 | 2 | NL | R96 | ✓ | U | ✓ | ✓ | - | $\Omega_{\mathrm{CP}}$ | -2.91 | -2.63 | -1.97 | 0.993 |
| CS11 | 2 | NL | R96 | ✓ | R | ✓ | ✓ | ✓ | $\Omega_{\mathrm{CP}}$ | -2.38 | -2.44 | -1.68 | 0.990 |
| CS12 | 2 | NL | R96 | ✓ | U | ✓ | ✓ | ✓ | $\Omega_{\mathrm{k}}$ | -1.70 | -1.90 | -1.39 | 1.005 |
| CS13 | 2 | L | R96 | ✓ | U | ✓ | ✓ | ✓ | $\Omega_{\mathrm{CP}}$ | -2.78 | 2.28 | -0.84 | 1.036 |
| CS14 | 2 | NL | R48 | ✓ | U | ✓ | ✓ | ✓ | $\Omega_{\mathrm{CP}}$ | -1.81 | -2.00 | -1.39 | 0.995 |

$g_O$: output activation function, Nonlinear (NL) as defined in Equation 24, or linear (L) i.e. $g_O(x) = x$
Data: uniform grid of 60x40 data points (U60x40), uniform grid of 96=12x8 data points (U12x8), and two randomly located sets of 96 (R96) and 48 (R48) data points.
CPs: uniform grid of 45x30 points (U), or random sampling of 45x30=1350 points at each epoch (R)
Auxiliary loss: calculated at CPs i.e. spatial domain $\Omega_{\mathrm{CP}}$, or in $k$-space $\Omega_k$ (see Equation 31)

Lowering the spatial density of data in CS7 (using a grid of test data U12x8) and further in CS8 (using randomly located test data R96) increases the reconstruction error of mode shapes, particularly at higher frequencies. As also compared in Figure 10(e,f), the elimination of the Gaussian mask in CS9 again slows down the convergence of k-PINN, and this time, it also leads to a significantly higher reconstruction error of mode shapes. This confirms the fact that the Gaussian mask plays the most crucial role when dealing with non-uniform and sparse data, which cannot provide sufficient information about the spatial distribution and $k$-space definition of (high-frequency) mode shapes.

Ignoring the sparsity loss term in case CS10 broadens the admissible $k$-space solution and slightly reduces the reconstruction error of mode shapes, while it affects the sparsity of the learnt $k$-space, which will be discussed later on. In CS11, instead of using a uniform grid of CPs, the same number of CPs is randomly resampled in every epoch. While such random resampling of CPs ensures satisfaction of governing PDEs at a multitude of spatial points, it can reduce the convergence rate of the algorithm. Here, it can be seen that random sampling of CPs slightly increases the reconstruction error of mode shapes compared to CS8, which used uniform sampling.

Case CS12 examines the effect of considering the $k$-space formulation of the auxiliary loss term, as defined in Equation 31. The results, also shown in Figure 10(g), clearly indicate the deficiency of this $k$-space formulation, which significantly increases the reconstruction error of mode shapes. This increase in error occurs because the principal output parameter (deflection) and the auxiliary parameters (its second order spatial derivatives) are directly and more strictly constrained in the $k$-space formulation, compared to when their $k$-space definition is independently inferred from the resultant deflection (after spectral superposition) and its derivatives. As such, the $k$-space formulation over-constrains the problem and leads to convergence to a suboptimal solution.

Given the relatively broadband frequency range of selected modes, the customized output activation function defined in Equation 24 was used for physics-informed cases thus far. As confirmed by case CS13 (Figure 10(h)), a linear output activation function fails in the reconstruction of several mode shapes from the higher frequency regime, while it leads to a slightly lower reconstruction error of the lower frequency mode shapes.

Lastly, case CS14 examines k-PINN by further reducing the density of data to half by using 48 randomly located data points (R48), leading to a higher reconstruction error compared to CS8 with 96 randomly located data points. Although the different case studies show varying performances in the reconstruction error of mode shapes, they all identify the stiffness scaling $\bar{\theta}_D \rightarrow 1$ with a relatively high and comparable accuracy, except in cases CS6 and CS9 in the absence of the Gaussian mask. This demonstrates that k-PINN is guiding the learning trajectory of all modes towards a valid (and reduced) solution in $k$-space, which satisfies the governing PDE with good accuracy.

A significant observation is the relatively fast and comparable convergence rate of all mode shapes with the application of the Gaussian mask in both data-driven (Figure 10(a)) and physics-informed (Figure 10(c,e, g)) cases, which clearly confirms the relieved spectral bias of the proposed $k$-space formulation. The results also show a faster convergence of the stiffness scaling $\bar{\theta}_D$, compared to the reconstruction of mode shapes (Figure 10(c,e)).

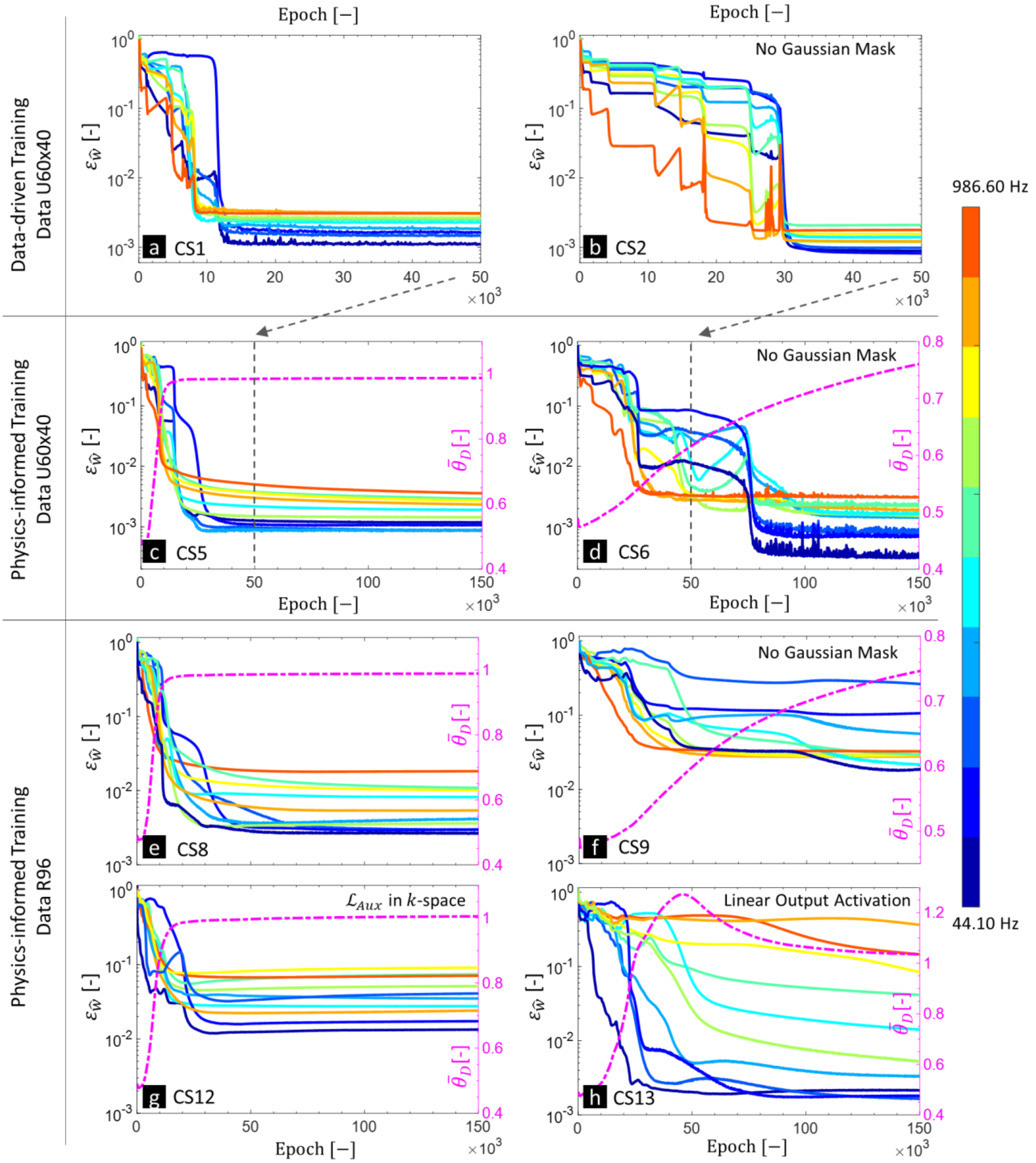

Figure 10. Evolution of (left axis) the reconstruction error and (right axis) the stiffness scaling $\bar{\theta}_D$ in function of training epochs, corresponding to the broadband selection of 10 modes with symmetric BCs (Figure 9). The panels correspond to selected cases listed in Table 3.

In order to gain better insight into the reconstruction accuracy of mode shapes, the selected modes S15, A46 and A72 are analyzed in more detail. Figure 11 displays the actual mode shapes with added white Gaussian noise (SNR=20 dB), as well as the reconstructed mode shapes using different spatial distributions of data points indicated with circular markers, corresponding to the cases CS7, CS8 and CS14. The results demonstrate the good performance of k-PINN in reconstruction of mode shapes of different frequencies using a relatively limited number of data points, particularly when the data points are randomly located, forming larger spatial gaps. Figure 12 further compares the full-field reconstruction error $\mathcal{E}_{\hat{w}_i}$ of selected modes, obtained from the different case studies. The left column shows the data-driven reconstruction error when using noise-free dense data U45x30, and the rest display the physics-informed reconstruction errors when using different cases of noisy data. A relatively low reconstruction error is achieved for the data-driven cases, due to the absence of physical constraints, but using a dense dataset. The physics-informed error naturally increases with the reduction of spatial density of data and further by considering randomly distributed data, particularly at higher frequency mode shapes. The reconstruction error is generally the highest close to the boundaries and more distinctively in the vicinity of supported corners.

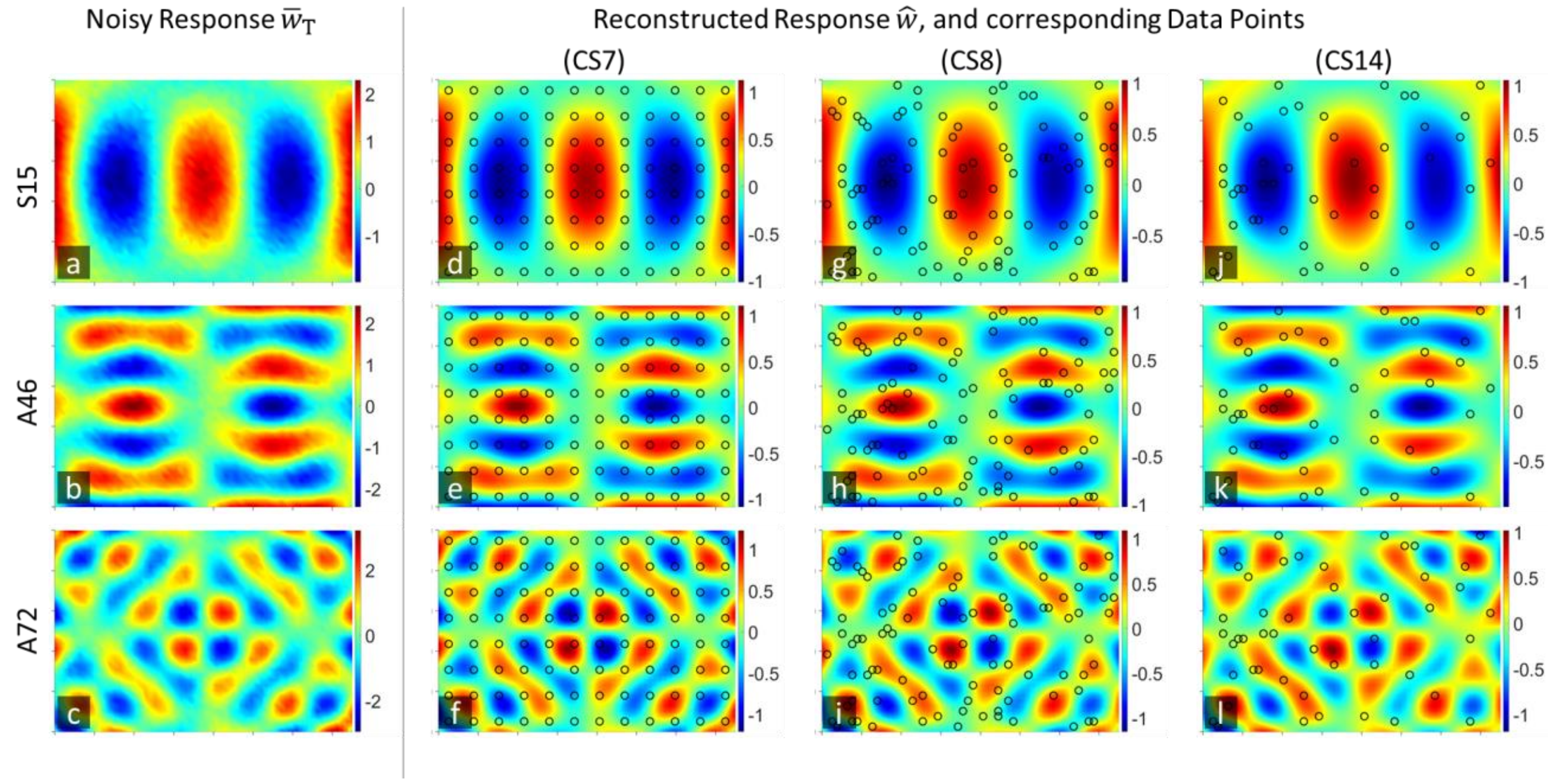

Figure 11. Performance of k-PINN for reconstruction of the bending deflection $w$ of selected modes S15, A46 and A72 for selected case studies CS7, CS8 and CS14 with different numbers and distributions of datapoints as listed in Table 3.

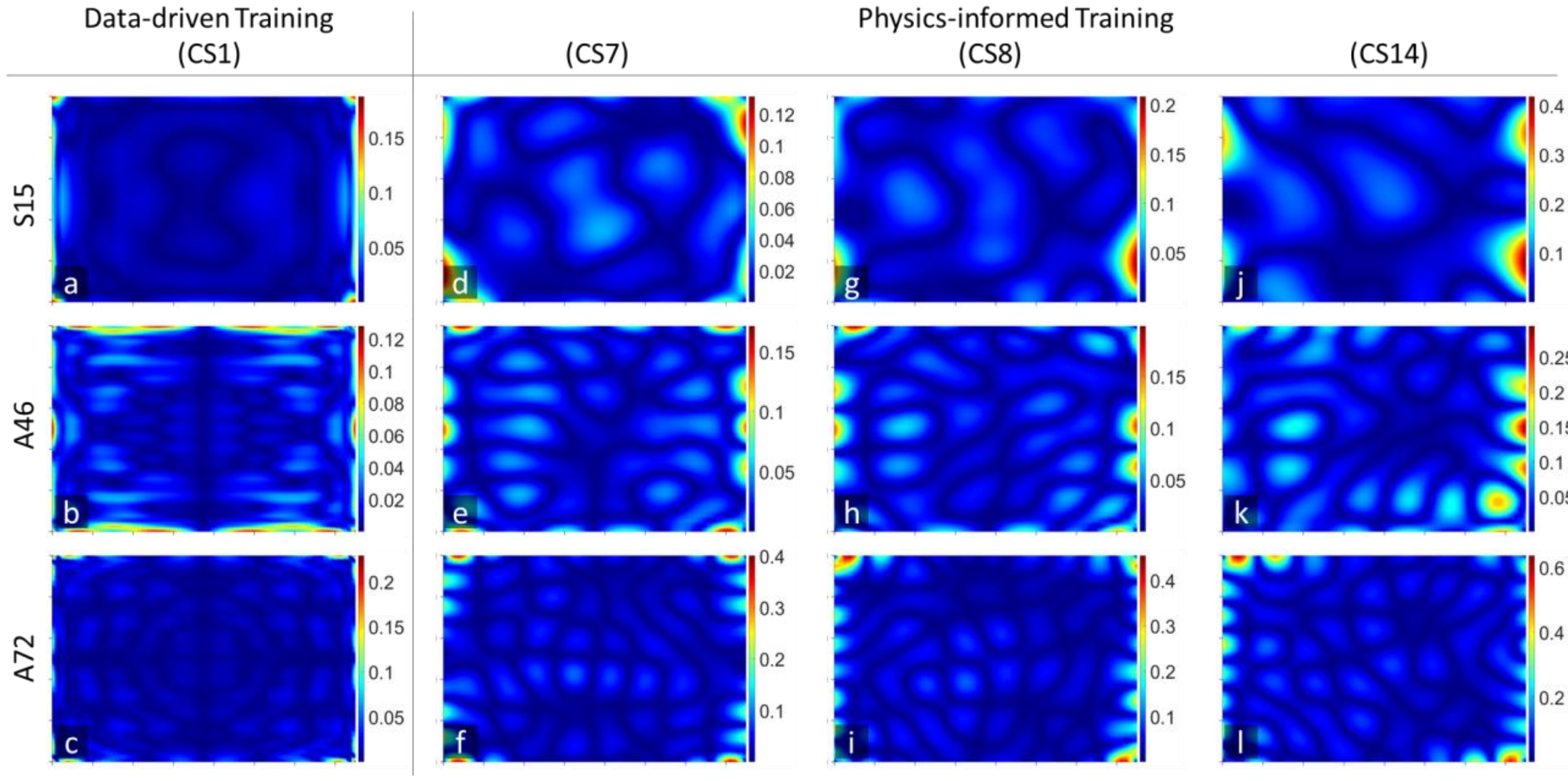

Figure 12. Performance of k-PINN in terms of the reconstruction error $\mathcal{E}_{\hat{w}_i}$ of selected modes S15, A46 and A72 for selected case studies: (a-c) data-driven case CS1, and (d-l) physics-informed cases CS7, CS8 and CS14, with different numbers and distributions of datapoints as listed in Table 3.

Furthermore, the $k$-space coefficient images of the three selected modes are shown in Figure 13 for case CS8 (with sparsity loss) and case CS10 (without sparsity loss). The results clearly confirm that application of the sparsity loss significantly compresses and organizes the $k$-space definition of all three mode shapes. Without the application of sparsity loss (Figure 13(a-f)), both real and imaginary components contribute comparably to the response, with a quite widespread distribution over the $k$-space. However, the sparsity loss compresses the $k$-space into a sparse set of spectral components (Figure 13(g-l)), and additionally provides a more physically representative definition of the mode shapes. The modes S15 and A72 (despite their opposite rectangular symmetry) are both diagonally symmetric and thus represented by a predominantly real $k$-space (Figure 13(g-h,k-l)), while mode A46 is diagonally anti-symmetric and thus represented by a predominantly imaginary $k$-space (Figure 13(i,j)). This is explained by the centralized spatial formulation introduced in section 3.2.1, which allows for the reconstruction of diagonally symmetric (anti-symmetric) modes by real (imaginary) $k$-space coefficients only.

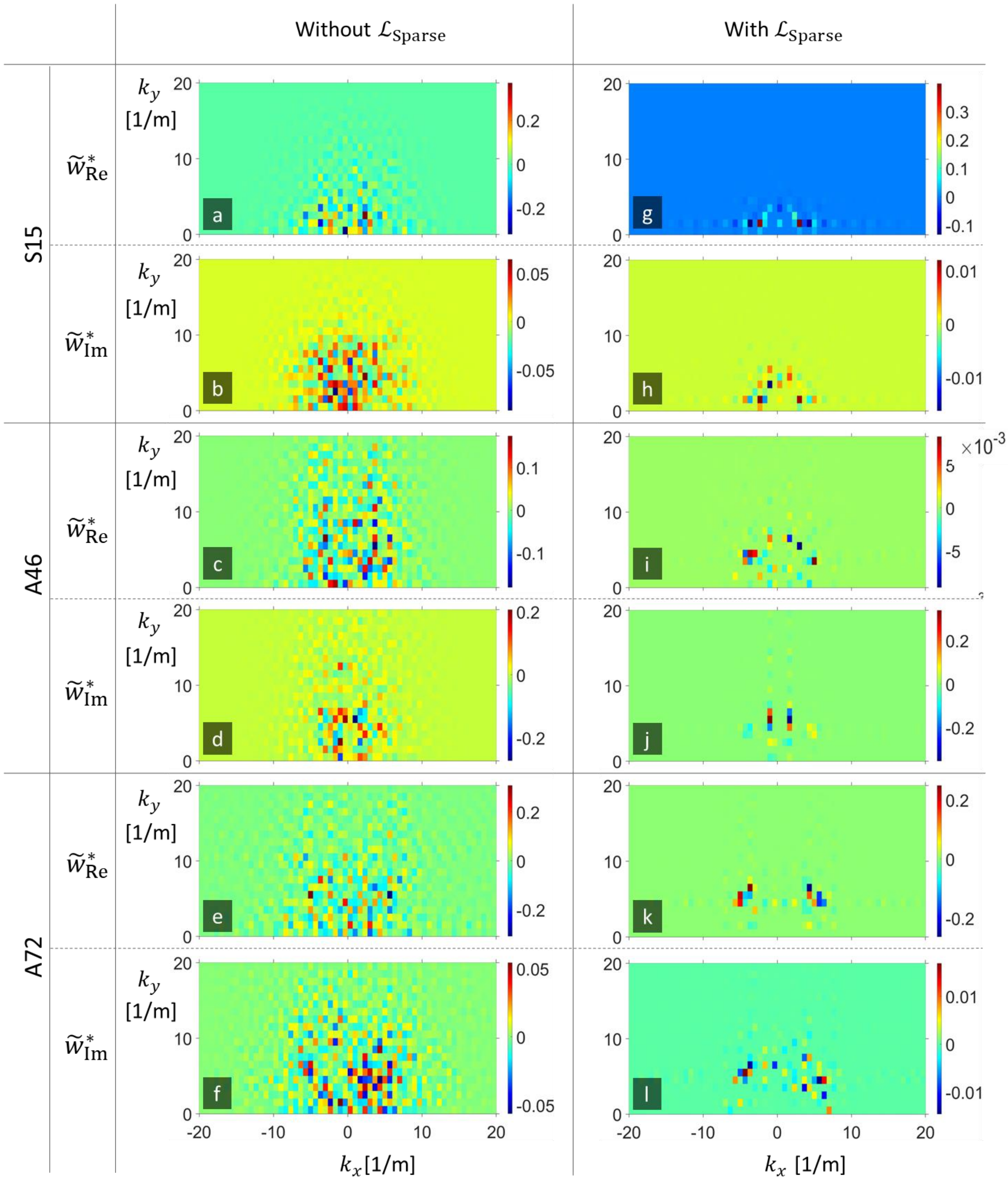

Figure 13. Real and imaginary coefficient images of the bending deflection, i.e. $\widetilde{w}^*_{\text{Re}}$ and $\widetilde{w}^*_{\text{Im}}$, corresponding to the selected mode shapes S15, A46 and A72, obtained from k-PINN, (left column) without and (right column) with considering the sparsity promoting loss, corresponding to cases CS8 and CS10 listed in Table 3.

#### *4.3.2. Asymmetric BCs*

In this section, k-PINN is further evaluated for learning a relatively broadband selection of 10 modes corresponding to the plate with asymmetric BCs, as shown in Figure 14. The aim is to demonstrate the performance of the proposed $k$-space formulation when dealing with asymmetric mode shapes. The mode numbers are chosen similar to those selected for the plate with symmetric BCs (Figure 9).

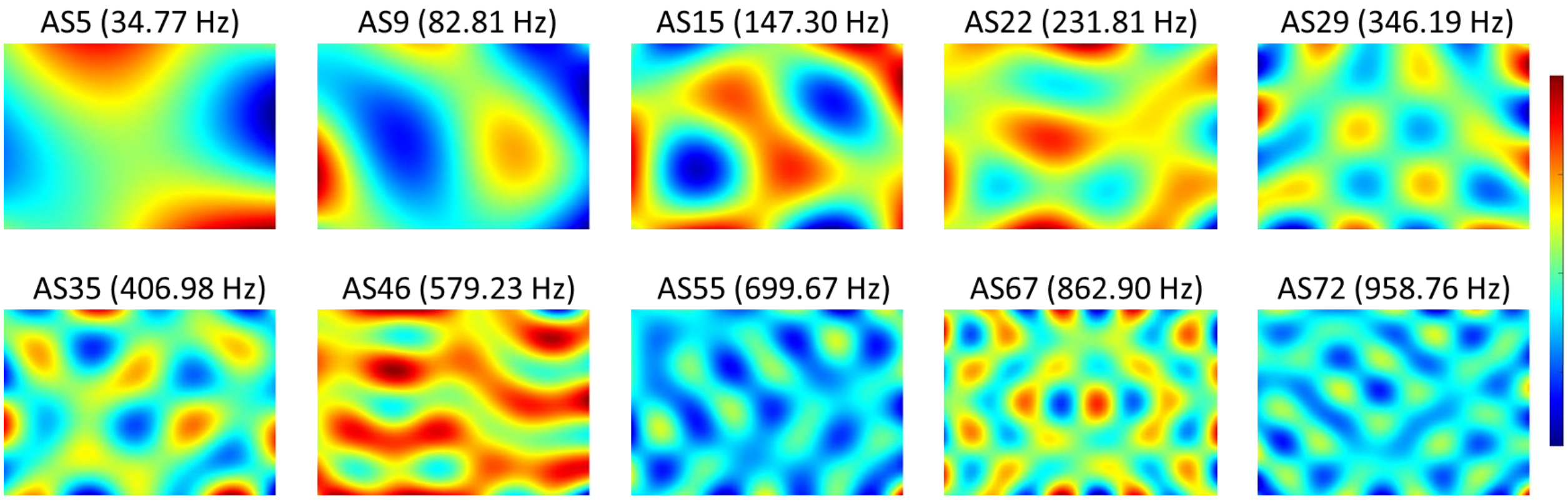

Figure 14. Selected broadband set of 10 modes for the plate with asymmetric BCs, i.e. free at the bottom-right corner and supported at the other three corners.

The hyperparameter settings and architecture of k-PINN is similar to those in the previous section. Six cases, denoted by CA1 to CA6, are studied as listed in Table 4, including the reconstruction error corresponding to the selected modes AS15, AS46 and AS72 and the identified stiffness scaling $\bar{\theta}_D$. The evolution of the reconstruction error for every individual mode as a function of the training epoch is further shown in Figure 15 for the physics-informed cases CA3 and CA5.

Overall, the results are in a good agreement with the previous section, with the main difference being that the reconstruction error of asymmetric modes is relatively higher than their symmetric counterparts. This can be attributed to the distinctive aperiodicity of mode shapes due to the presence of a free corner, leading to a broadband $k$-space representation that may not fit within the bandwidth prescribed for k-PINN.

Nonetheless, similar to the case of symmetric modes, the stiffness scaling is inferred with relatively high accuracy ($\bar{\theta}_D \rightarrow 1$), meaning that the underlying PDE is satisfied with good accuracy. This again confirms that k-PINN reconstructs an optimized representation of mode shapes with respect to the prescribed $k$-space solution space, given the test data and the 4th order smoothness imposed by the governing PDE.

Table 4. Performance of k-PINN for simultaneously learning the broadband selection of 10 modes with asymmetric BCs (Figure 14). All NNs have five hidden layers and a constant width of 128.

| Case ID | NN settings | | Data | Noise | CPs | k-PINN settings | | | | $\log_{10} \varepsilon_{\hat{w}}$ | | | $\bar{\theta}_D$ |
|---|---|---|---|---|---|---|---|---|---|---|---|---|---|
| | $N_{\mathrm{NN}}$ | $g_O$ | | | | Phys. Loss | Gauss. Mask | Spars. Loss | Aux. Loss | S15 | A46 | A72 | |
| CA1 | 1 | NL | U60x40 | - | U | - | ✓ | ✓ | $\Omega_{\mathrm{CP}}$ | -2.70 | -2.61 | -2.46 | - |
| CA2 | 1 | NL | U60x40 | - | U | - | - | ✓ | $\Omega_{\mathrm{CP}}$ | -3.15 | -2.72 | -2.42 | - |
| CA3 | 2 | NL | U60x40 | ✓ | U | ✓ | ✓ | ✓ | $\Omega_{\mathrm{CP}}$ | -2.83 | -2.67 | -2.18 | 0.968 |
| CA4 | 2 | NL | U12x8 | ✓ | U | ✓ | ✓ | ✓ | $\Omega_{\mathrm{CP}}$ | -2.35 | -2.07 | -1.47 | 0.962 |
| CA5 | 2 | NL | R96 | ✓ | U | ✓ | ✓ | ✓ | $\Omega_{\mathrm{CP}}$ | -2.09 | -1.57 | -1.31 | 0.983 |
| CA6 | 2 | NL | R48 | ✓ | U | ✓ | ✓ | ✓ | $\Omega_{\mathrm{CP}}$ | -1.39 | -1.29 | -1.15 | 0.978 |

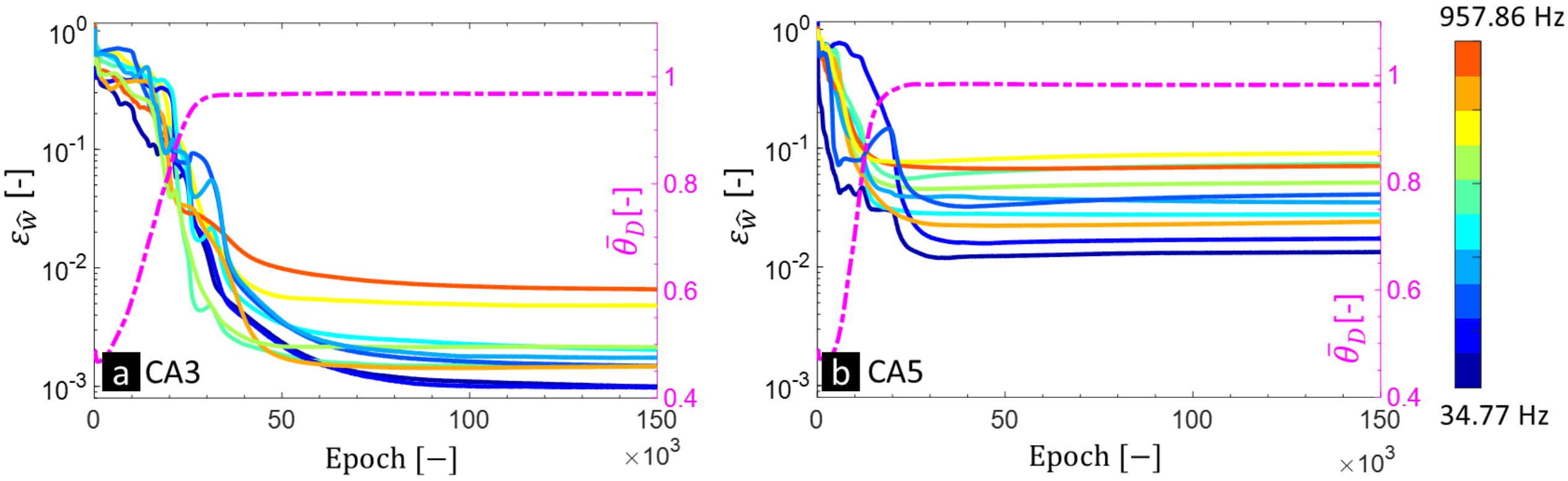

Figure 15. Evolution of (left axis) the reconstruction error and (right axis) the stiffness scaling $\bar{\theta}_D$ in function of training epochs, corresponding to the broadband selection of 10 modes with asymmetric BCs (Figure 14), for physics-informed cases: (a) CA3 with a dense uniform grid of data U45x30 and (d) CA5 with randomly located data R96.

Figure 16 compares the selected mode shapes after addition of noise with the mode shapes reconstructed by k-PINN, with corresponding data points indicated by circular markers. The results confirm the good correlation between the simulated and reconstructed mode shapes at different frequencies, using limited and randomly distributed data.

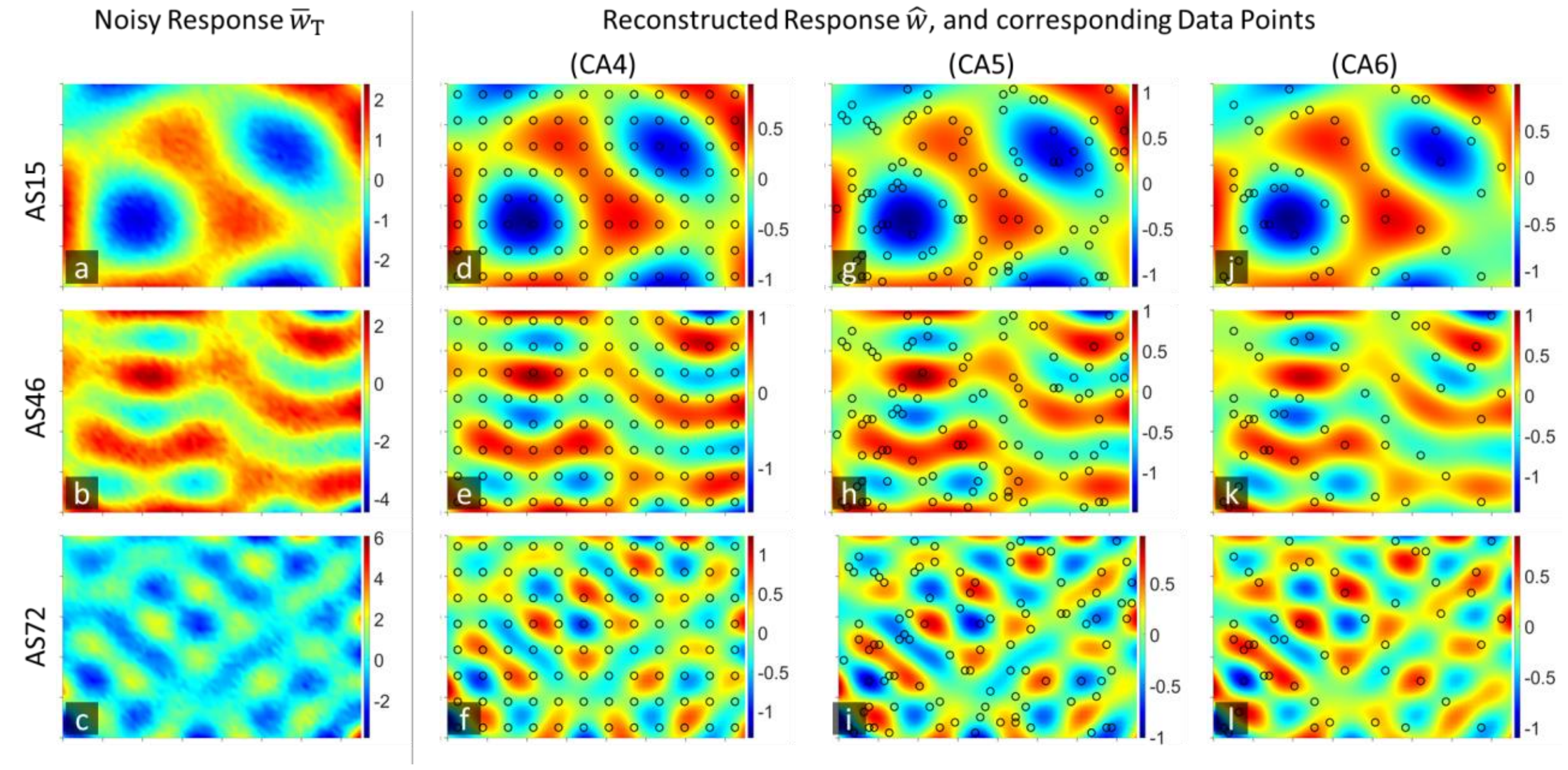

Figure 16. Performance of k-PINN for reconstruction of the bending deflection $w$ of selected asymmetric modes AS15, AS46 and AS72. Circular markers indicate the location of data points used for reconstruction.

## 4.4. K-PINN for a Narrowband Selection of Mode Shapes and Identification of Individual Bending Stiffness Components

In all the experiments discussed in the previous sub-sections, k-PINN was applied for inferring an overall stiffness scaling factor $\bar{\theta}_D$ as the only unknown stiffness parameter, with a focus on the reconstruction of a wideband selection of modes. In this section, a relatively narrowband selection of 10 modes are picked from the low frequency regime, to identify the individual stiffness scaling factors. The two selected sets of mode shapes corresponding to the symmetric and the asymmetric BCs are shown in Figure 17.

From the governing equation of motion (Equation 7), it can be seen that the twisting component of bending motion is governed by a combinatory stiffness $D_C = (D_{12} + 2D_{66})$ over the entire surface of the plate. As such, the individual stiffness components $D_{12}$ and $D_{66}$ can only be identified through an accurate satisfaction of the Kirchhoff free edge condition (Equation 9), which is governed by another combinatory stiffness $(D_{12} + 4D_{66}) = D_C + 2D_{66}$. Unfavourably, as shown in Figure 12, the reconstruction error is the highest at the boundaries, a known challenge in PINNs [48; 49]. Moreover, the two stiffness components $D_{12}$ and $D_{66}$ have a relatively very small magnitude compared to the stiffness components $D_{11}$ and $D_{22}$ (Table 1), and thus contribute proportionally less to the governing PDE. All these factors lead to the conclusion that individual identification of the stiffness components $D_{12}$ and $D_{66}$ from this vibrational case study is not reliable. It is worth noting that, the shape of the vibrating plate, its boundary conditions and the selected set of mode shapes all can be further optimized for maximized sensitivity and faster convergence in identification of material properties [50], which is not the focus of this study.

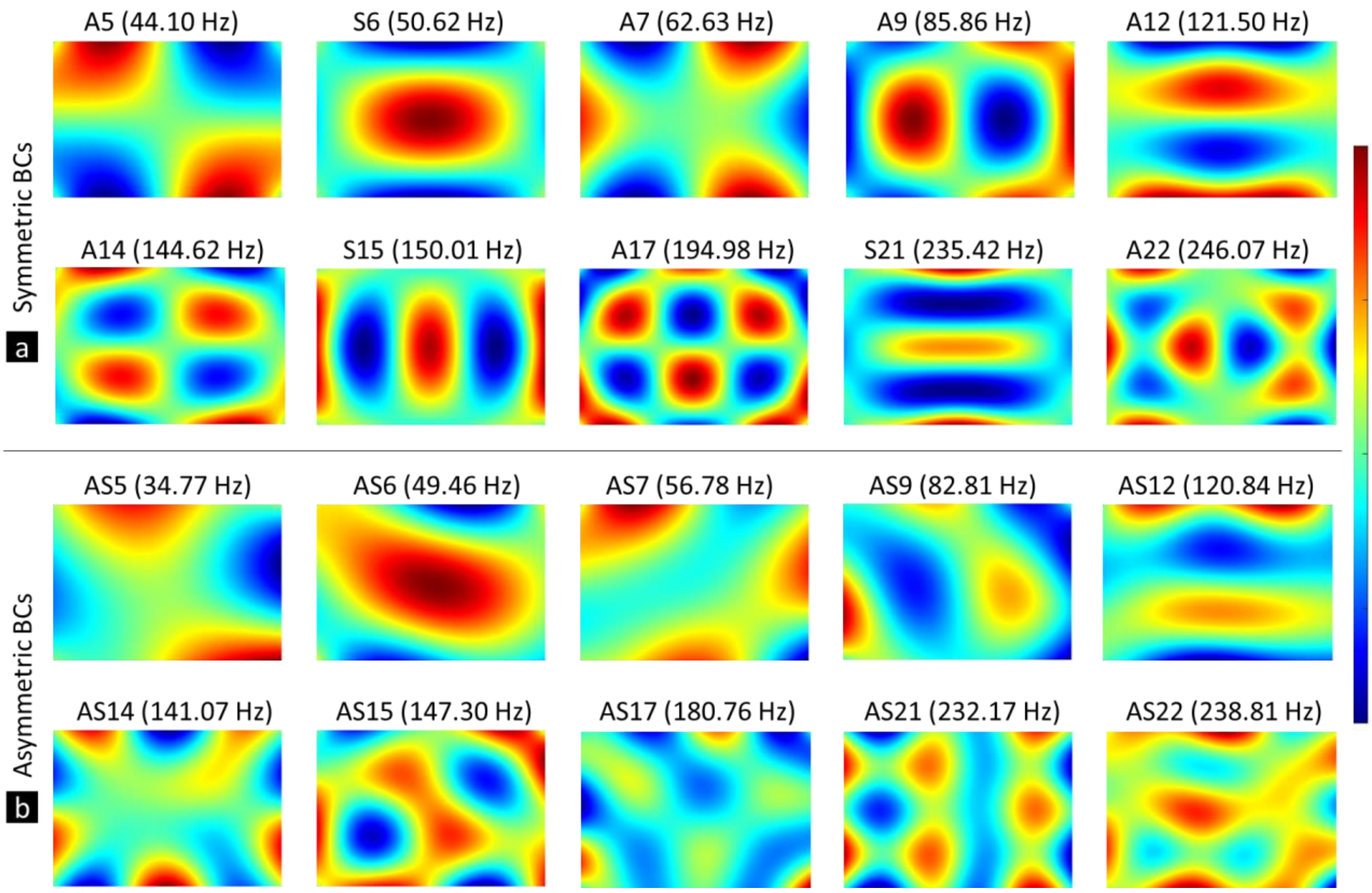

Figure 17. Selected narrowband set of 10 modes: (a) for the plate with symmetric BCs supported at the four corners, and (b) for the plate with asymmetric BCs i.e. free at the bottom-right corner and supported at the other three corners.

Hence, a fixed stiffness scaling $\theta_{12} = 1$ is considered, and the problem is reduced to the identification of the remaining three components $\theta_{11}$, $\theta_{22}$ and $\theta_{66}$, from disproportional initial guess values of 0.2, 0.3 and 0.4, respectively. Given the relatively low frequency range of selected modes, a linear output activation function is used, which, according to the results of section 4.3.1 (Figure 10(e,h)), showed slightly lower reconstruction error for the low-frequency mode shapes. The applied k-PINN settings are summarized in Table 5.

Table 5. Settings of k-PINN for simultaneously learning the narrowband selections of 10 modes (Figure 17). All NNs have five hidden layers and a constant width of 128.

| NN settings | | Data | Noise | CPs | k-PINN settings | | | |
|---|---|---|---|---|---|---|---|---|
| $N_{\mathrm{NN}}$ | $g_O$ | | | | Phys. Loss | Gauss. Mask | Spars. Loss | Aux. Loss |
| 2 | L | U12x8 | ✓ | U | ✓ | ✓ | ✓ | $\Omega_{\mathrm{CP}}$ |

Figure 18 shows the evolution of modal reconstruction errors and the identification of stiffness scaling factors as a function of training epochs. Identification is performed with five initializations of k-PINN, and the relevant means and standard deviations are respectively shown with dashed lines and shaded areas in Figure 18(b,d).

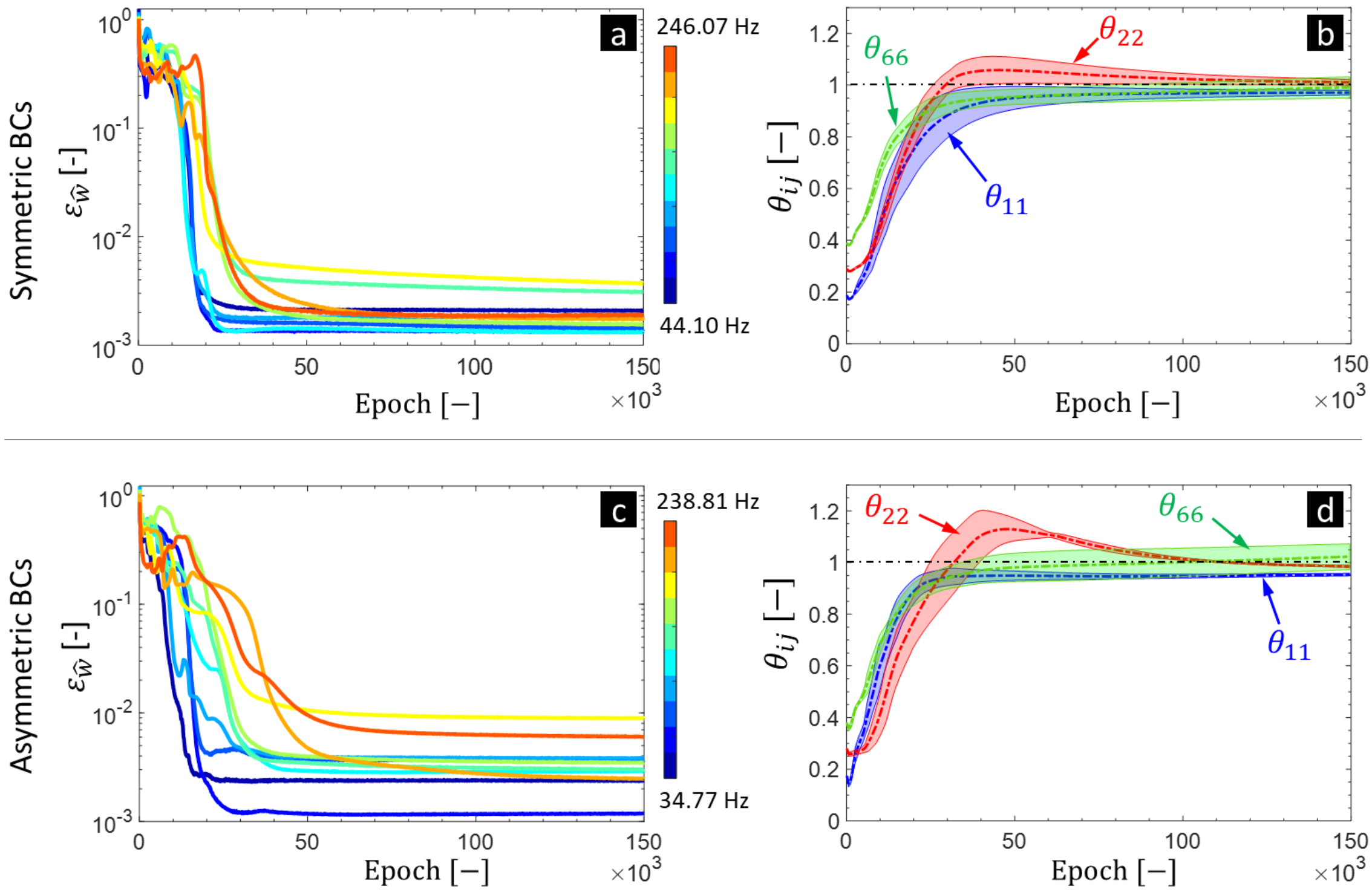

Figure 18. Evolution of (a,c) the reconstruction error of the narrowband selection of 10 modes (Figure 17), and (b,d) mean (dashed line) and standard deviation (shaded area) of stiffness scaling parameters from five initializations of k-PINN, in function of training epochs.

According to the results, the modes corresponding to the asymmetric BCs experience a generally lower convergence rate and a slightly higher reconstruction error (Figure 18(c)). As such, the stiffness parameters identified from the modes with symmetric BCs (Figure 18(b)) show slightly higher accuracy compared to those identified from asymmetric modes (Figure 18(d)). Nonetheless, the final parameters identified from both cases, as listed in Table 6, show good accuracy close to the true value of unity.

Table 6. Identified bending stiffness scaling parameters from the narrowband selection of 10 modes for symmetric and asymmetric BCs. All stiffness scaling parameters have a true value of unity.

| | | Bending Stiffness scaling factor [−] | | | |
|---|---|---|---|---|---|
| | | $\theta_{D11}$ | $\theta_{D22}$ | $\theta_{D66}$ | $\theta_{D12}$ (Constant) |
| Initial Guess | | 0.2 | 0.3 | 0.4 | 1.0 |
| Identified | Sym. BCs | 0.971±0.009 | 1.011±0.011 | 0.991±0.041 | 1.0 |
| | Asym. BCs | 0.953±0.005 | 0.985±0.039 | 1.023±0.050 | 1.0 |

### 4.5. Computational efficiency

Because of the spectral mapping of the response function in k-PINN, and the relaxation of the need for additional sequential backpropagations through NNs for the generation of PDE terms and computation of physics-informed loss values,  higher computational efficiency than that of standard PINN is expected. To verify this, the computational costs of PINN and k-PINN are compared by computing their processing time per epoch, averaged over 1000 epochs. Computations are performed on a GPU, using a machine with Intel(R) Core(TM) i7-12700H  2.30 GHz, installed RAM memory of 32 GB, and an available GPU of 15.8 GB.

Figure 19(a,b) compares the computational cost of the two algorithms with respect to the width of NNs and number of CPs, while Figure 19(c,d) compares the computational cost of k-PINN for different numbers of spectral components in $k$-space and different NN structures considered in section 4.3. By default, a NN width of 128, number of CPs of $45 \times 30 = 1350$, and a number of spectral components of 1281 are considered, corresponding to the defaults settings applied in this study for reconstruction of mode shapes.

The results confirm the consistently lower computational time of k-PNN, which is significantly more pronounced when a wider NN is used (Figure 19(a)) and when more CPs are considered (Figure 19(b)). It can be seen that, contrary to PINN, the computational speed of k-PINN is almost insensitive to the width of NNs. This can be explained by the relaxation of additional backpropagations for the calculation of physics-informed loss terms. The lower computational cost of k-PINN, with respect to the number of CPs, can be attributed to its spectral mapping, which generates the entire spatial response from corresponding $k$-space coefficients, obtained with a single feed forward solution of NNs.

The computational cost of k-PINN itself can vary with the size (bandwidth and/or resolution) of $k$-space and its NN's architecture. Figure 19(c) shows that the computational time disproportionally changes by varying the number of $k$-space components. Increasing the number of components by a factor of four (e.g. doubling the bandwidth or resolution) increases the computational time by around 50%. Reducing the number of components, however, only slightly reduces the computational time.

To further reduce the computational cost of k-PINN, particularly for a larger number of CPs and larger $k$-space sizes, one may perform the spectral mapping much more efficiently using 2D fast inverse Fourier transform. This would merely limit the spatial distribution of measurement points and CPs to a prescribed uniform spatial grid. It is also noteworthy that the higher computational efficiency of k-PINN is more pronounced if the computations are performed on CPU instead of GPU.

Figure 19(c) shows the computational time of k-PINN with one and two NNs and their minor difference, corresponding to the two different architectures examined in sections 4.3 and 4.4.

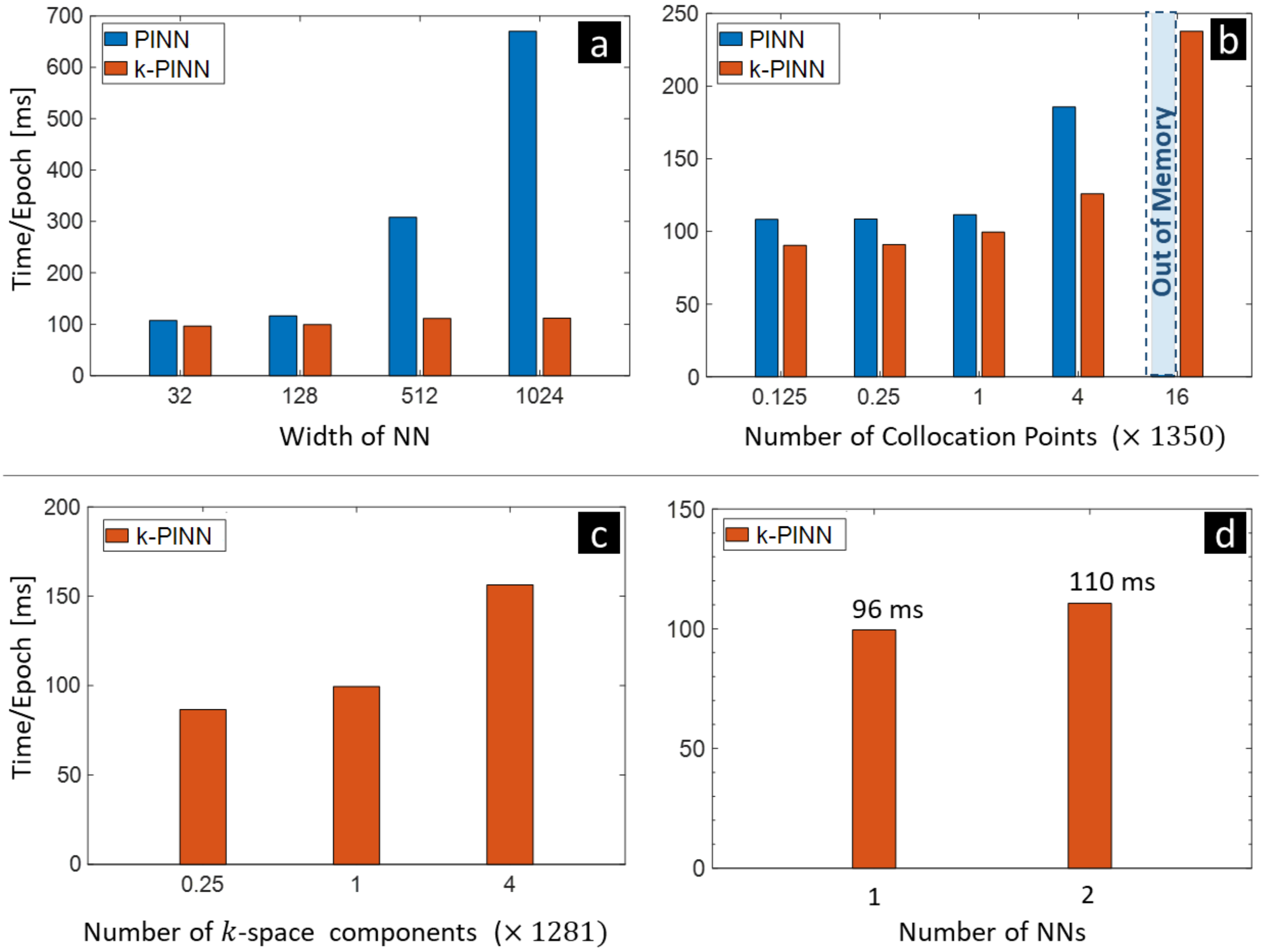

Figure 19. Computational cost of PINN and k-PINN in terms of processing time per each epoch of training (herein the mean value of 1000 epochs), with respect to: (a) width of NN, (b) Number of collocation points (CPs), (c) number of spectral components in $k$-space and (d) number of NNs. By default, a NN width of 128, number of CPs of $45 \times 30 = 1350$, and number of spectral components of 1281, are considered.

## 5. Conclusions

In this research, in view of the efficient reconstruction of broadband vibrational mode shapes from sparse measurement data, a spectral formulation of physics-informed neural networks (PINNs) was introduced, which maps the response function in $k$-space. The proposed $k$-space PINN (k-PINN), thanks to its spectral formulation: (i) provides a rigorous solution space for the reconstruction of broadband vibrational mode shapes with multiscale spatial features, and (ii) enables efficient computation of the response function and its spatial derivatives of any order at once, by relaxing the need for additional backpropagations throughout the neural network for the calculation of the governing equation. k-PINN was evaluated for the reconstruction of bending vibrational mode shapes of a thin composite laminate and the identification of its orthotropic bending stiffness components, from a virtual test dataset with added white Gaussian noise.

Comparison of k-PINN with standard spatial PINN confirmed both its superior performance in the reconstruction of a single vibrational mode shape and its significantly lower computational cost. k-PINN was further evaluated for simultaneously reconstructing different selections of 10 mode shapes, up to 1000 Hz, and the identification of bending stiffness coefficients. Symmetric (asymmetric) boundary conditions were examined for the reconstruction of corresponding symmetric/anti-symmetric (asymmetric) mode shapes. Both data-driven and physics-informed scenarios were studied, demonstrating the high performance of proposed k-PINN in terms of a relatively fast learning and reconstruction of all mode shapes with a comparable convergence rate (i.e. relaxed spectral bias).

The results showed that the reconstruction error increases with the sparsity of input measurement data and the strong aperiodicity induced by asymmetric boundary conditions. Nonetheless, it was shown that k-PINN consistently converges to a valid, and reduced $k$-space solution, which has a high correlation with the exact solution, satisfies the governing equation, and identifies the bending stiffnesses with good accuracy.

It is possible to define the complexity of the response function, or perform model order reduction, by adjusting the bandwidth and/or resolution of the deterministic $k$-space prescribed for k-PINN. Potential future research directions include enhancing the reconstruction accuracy and computational efficiency of k-PINN through efficient exploration of the sparse $k$-space, and further through a transfer learning approach (i.e. pre-training of k-PINN with a reference dense simulation/measurement data, and then fine-tuning it with sparse test data at deployment). Moreover, k-PINN can be further expanded to spatio-temporal reconstruction of transient vibrations, by additionally incorporating a spectral formulation of the time dimension. These steps can pave the ground for the application of k-PINN to more complex physical problems, including nonlinear vibrations and spatially varying material properties.

In principle, the proposed k-PINN can be applied to any physical problem, particularly when the spectral sparsity is more prominent like the vibrational case study evaluated in this work. When dealing with other problems, the bandwidth and resolution of prescribed $k$-space, the regularization mask, and the definition of auxiliary parameters and relevant settings have to be reconsidered and adapted to ensure the efficient convergence and reconstruction accuracy of k-PINN.

**Appendix**

The in-plane orthotropic elastic constants $C_{ij}$ of a unidirectional fibrous composite ply can be derived from its engineering elastic modulus as follows:

$$\begin{bmatrix} C_{11} & C_{12} & 0 \\ C_{12} & C_{22} & 0 \\ 0 & 0 & C_{66} \end{bmatrix} = \begin{bmatrix} \dfrac{E_1}{1-\nu_{12}\nu_{21}} & \dfrac{\nu_{12}E_1}{1-\nu_{12}\nu_{21}} & 0 \\ \dfrac{\nu_{12}E_1}{1-\nu_{12}\nu_{21}} & \dfrac{E_2}{1-\nu_{12}\nu_{21}} & 0 \\ 0 & 0 & G_{12} \end{bmatrix} \quad (36)$$

where $E_1$ and $E_2$ are the elastic moduli parallel and perpendicular to the fibers, $G_{12}$ is the in-plane shear modulus, and $\nu_{12} = (E_1/E_2)\nu_{21}$ is the in-plane Poisson's ratio. Subsequently, the effective elastic constants of a ply with an orientation angle $\theta$ with respect to the $x$-axis can be derived as:

$$\begin{cases} \bar{C}_{11} = C_{11}\cos^4\theta + 2(C_{12} + 2C_{66})\sin^2\theta\cos^2\theta + C_{22}\sin^4\theta \\ \bar{C}_{12} = (C_{11} + C_{22} - 4C_{66})\sin^2\theta\cos^2\theta + C_{12}(\sin^4\theta + \cos^4\theta) \\ \bar{C}_{22} = C_{11}\sin^4\theta + 2(C_{12} + 2C_{66})\sin^2\theta\cos^2\theta + C_{22}\cos^4\theta \\ \bar{C}_{16} = (C_{11} - C_{12} - 2C_{66})\sin\theta\cos^3\theta + (C_{12} - C_{22} + 2C_{66})\sin^3\theta\cos\theta \\ \bar{C}_{26} = (C_{11} - C_{12} - 2C_{66})\sin^3\theta\cos\theta + (C_{12} - C_{22} + 2C_{66})\sin\theta\cos^3\theta \\ \bar{C}_{66} = (C_{11} + C_{22} - 2C_{12} - 2C_{66})\sin^2\theta\cos^2\theta + C_{66}(\sin^4\theta + \cos^4\theta) \end{cases} \quad (37)$$

Given the effective elastic constants $\bar{C}_{ij}$ of each ply, the effective bending stiffness coefficients corresponding to a composite laminate with $N$ identical plies of thickness $h_p$, are defined as:

$$D_{ij} = \frac{1}{3}\sum_{n=1}^{N} \overline{C}_{ij}^{(n)}\left(z_{n+1}^3 - z_n^3\right) \quad (38)$$

where $z_n = (n - 1 - 0.5N)h_p$ is the coordinate of the bottom face of ply $n$ (ordered from bottom to top) with respect to the mid-plane of the laminate. For the cross-ply laminate considered in this study, we have $D_{16} = D_{26} = 0$.

**Acknowledgement**

The authors acknowledge funding from Bijzonder Onderzoeksfonds (BOF) of UGent through grant (BOF21/PDO/041), and Fonds voor Wetenschappelijk Onderzoek Vlaanderen (FWO-Vlaanderen) through grant V430922N.